\newif\ifpdflatex    % toggle between pdf and ps images
\pdflatextrue           % use with pdflatex (plat.sh)
\documentclass[usenatbib]{mn2e}
\usepackage{graphicx}
\usepackage{soul}
\def\lesssim{\mathrel{\hbox{\rlap{\hbox{\lower5pt\hbox{$\sim$}}}\hbox{$<$}}}}
\def\gtrsim{\mathrel{\hbox{\rlap{\hbox{\lower5pt\hbox{$\sim$}}}\hbox{$>$}}}}
\title[SN 2008S \& N300OT are likely SNe]{Almost Gone: SN~2008S and NGC~300 2008OT-1 are Fainter than their Progenitors}
\author[Adams et al.]
  {\parbox{18cm}{S.~M.~Adams$^{1,2}$, C.~S.~Kochanek$^{1,2}$, J.~L.~Prieto$^{3,4}$, X.~Dai$^{5}$, B.~J.~Shappee$^{6,7,8}$, and K.~Z.~Stanek$^{1,2}$}
  \\
  \\
  $^{1}$ Dept.\ of Astronomy, The Ohio State University, 140 W.\ 18th   Ave., Columbus, OH 43210\\
  $^{2}$ Center for Cosmology and AstroParticle Physics (CCAPP), The Ohio State University, 191 W.\ Woodruff Ave., Columbus, OH 43210\\
  $^{3}$ N\'ucleo de Astronom\'ia de la Facultad de Ingenier\'ia, Universidad Diego Portales, Av. Ej\'ercito 441, Santiago, Chile \\
  $^{4}$ Millennium Institute of Astrophysics \\
  $^{5}$ Department of Physics and Astronomy, University of Oklahoma, 440 W.\ Brooks St., Norman, OK 73019, USA \\
  $^{6}$ Carnegie Observatories, 813 Santa Barbara Street, Pasadena, CA 91101, USA \\
  $^{7}$ Carnegie-Princeton Fellow \\
  $^{8}$ Hubble Fellow \\
  E-mail: sadams@astronomy.ohio-state.edu}
\begin{document}
\voffset -1.5cm
\maketitle

\begin{abstract}
We present late-time \emph{Hubble} and \emph{Spitzer Space Telescope} imaging of SN 2008S and NGC 300 2008OT-1, the prototypes of a common class of stellar transients whose true nature is debated.  
Both objects are still fading and are now $>15$ times fainter than the progenitors in the mid-IR and are undetected in the optical and near-IR.
Data from the Large Binocular Telescope and Magellan show that neither source has been variable in the optical since fading in 2010.
We present models of surviving sources obscured by dusty shells or winds and find that extreme dust models are needed for surviving stars to be successfully hidden by dust.
Explaining these transients as supernovae explosions, such as the electron capture supernovae believed to be associated with extreme AGB stars, seems an equally viable solution.
Though SN 2008S is not detected in \emph{Chandra X-Ray Observatory} data taken in 2012, the flux limits allow the fading IR source to be powered solely by the shock interaction of ejecta with the circumstellar medium if the shock velocity at the time of the observation was $\gtrsim20\%$ slower than estimated from emission line widths while the transient was still optically bright.
Continued \emph{SST} monitoring and $10-20\>\mu\mathrm{m}$ observations with \emph{JWST} can resolve any remaining ambiguities.
\\
\\
\end{abstract}

%\begin{keywords}
%stars: evolution -- supergiants -- supernovae: general -- supernovae: individual (NGC 300-OT)
%\end{keywords}

\section{Introduction}
SN 2008S-like events are transients arising from heavily obscured extreme asymptotic branch stars \citep{Prieto08,Thompson09}.  The spectra of the events are similar to Type IIn supernovae (SNe), but have lower ejecta velocities ($\sim1000$ km/s) and peak luminosities ($\sim -10$ to $-15$ mag) than typical supernovae \citep[e.g.,][]{Smith11}.

SN 2008S-like transients are not a rare, inconsequential phenomenon.  Though few events have been detected due to their low luminosities, the rate of these transients is $\sim10-20\%$ of the core-collapse supernova (ccSN) rate \citep{Thompson09}.  The obscured progenitors of these transients are very rare \citep[even relative to massive stars;][]{Thompson09,Khan10}, which likely means that the dust-enshrouded phase is a relatively common but short-lived ($<10^{4}$ yr) phase \citep{Thompson09}.

These transients are often considered to be a subclass of SN impostors.  However, other SN impostors seem to arise from more massive stars ($>20~M_{\odot}$) and are often considered to be eruptions of Luminous Blue Variables \citep[see, e.g.,][]{Humphreys94,Smith11,Kochanek12}.  While some events classified as SN impostors clearly are non-terminal, evidence is emerging that others are just as likely to be low-luminosity, core-collapse events \citep{Kochanek12,Adams15}.  

The two best prototypes of the SN 2008S class are SN 2008S itself \citep{Arbour08} and the very similar NGC 300 2008OT-1 \citep[][hereafter referred to as N300OT]{Monard08}.
The progenitor of SN 2008S was heavily obscured and undetected in the optical, but was identified as a mid-IR source with $L_{*}\simeq 10^{4.5}~L_{\odot}$ and a blackbody temperature of $T_{\mathrm{bb}}\simeq 440$ K \citep[see Table \ref{tab:prog08s};][]{Prieto08}.  The transient peaked at an absolute V-band magnitude of $-14.0 \pm 0.2$ \citep{Botticella09}.  Likewise, the dusty progenitor of N300OT had a luminosity of $10^{4.9}~L_{\odot}$ and $T_{\mathrm{bb}}\simeq 300$ K \citep[see Table \ref{tab:n300prog};][]{Prieto08atel}.  The N300OT transient peaked at $-12.9 < M_{V} < -12.0$ depending on the reddening estimate \citep{Bond09}.

\cite{Kochanek11b} concluded that SN 2008S and N300OT were explosive transients where a radiation spike occurring when shocks broke out from the surface of the star temporarily destroyed their encasing dust cocoons.
\cite{Botticella09} supports the interpretation that SN 2008S is a weak electron-capture supernova (ecSN) of a super-asymptotic giant branch progenitor, finding the quasi-bolometric light out to 300 days to be consistent with the decay of $^{56}$Co.  However, \cite{Smith09} find a substantial bolometric correction at 270 days that makes the true decay rate (0.06 mag day$^{-1}$) almost half that of $^{56}$Co.  Moreover, they find the spectrum to be similar to that of a Galactic hypergiant, and favor interpreting the event as the cool super-Eddington wind of an LBV in eruption.  Likewise, \cite{Berger09} favor a similar non-terminal event for N300OT on the basis of its low energy and spectroscopic similarities to an active yellow hypergiant.  Meanwhile, \cite{Kashi10} propose that N300OT was the result of mass transfer episode from an extreme asymptotic giant branch star to a main-sequence companion.  \cite{Prieto09} find similarities between the spectrum of N300OT and proto-planetary nebulae consistent with a $\sim 6-10 M_{\odot}$ carbon-rich asymptotic giant branch (AGB), super-AGB, or post-AGB star as the progenitor.

Ultimately, whether these transients were terminal events must be settled by late-time imaging to see if the sources either vanish or settle back to luminosities similar to that of the progenitors.  \cite{Prieto10} and \cite{Szczygiel12} found SN 2008S to be brighter than the progenitor but still fading in 2010 and 2011.  
While the evolution of other, older SN 2008-like transients such as SN 1999bw and SN 2002bu have also been followed and show the sources to be fading \citep{Kochanek12,Szczygiel12b}, these test cases are less powerful because there were no pre-transient detections of the progenitors.

We have continued to monitor the prototypes of this class of transients, SN 2008S and N300OT, with the \emph{Hubble Space Telescope} (\emph{HST}), the \emph{Spitzer Space Telescope} (\emph{SST}), the \emph{Chandra X-ray Observatory}, the Large Binocular Telescope (LBT) and Magellan.  In \S2 we present late-time data showing that these objects have faded below the luminosities of their progenitors.  We then introduce the methods and models we use to constrain the existence of any surviving stars.  In \S3 we present the results of modeling the spectral energy distributions (SEDs) of the sources and evaluate whether surviving stars could be hidden behind dust.  In \S4 we summarize the results and discuss the implications.

We adopt distances of 1.88 Mpc to NGC 300 \citep{Gieren05} and 5.6 Mpc to NGC 6946 \citep{Sahu06} and Galactic foreground extinctions of $E(B-V) = 0.011$ for NGC 300 and 0.303 mag for NGC 6946 based on the \cite{Schlafly11} recalibration of the \cite{Schlegel98}.

\section{Data and Models}

\subsection{Data}

\subsubsection{Optical and IR}

We utilize both new and archival \emph{SST} data.  We obtained a series of 3.6 and 4.5$~\mu\mathrm{m}$ images between 2010 and 2015 (program IDs 70040, 80015, 90124, 10081, and 11084).  We supplemented this data with available archival images (program IDs: 159, 1083, 3248, 10136, 20256, 20320, 30292, 30494, 40010, 40204, 40619, 61002, 80196, 90178; PIs: J. Andrews, M. Barlow, W. Freedman, G. Helou, M. Kasliwal, R. Kennicutt, R. Kotak, W.P. Meikle, M. Meixner, B. Sugerman).
We followed the evolution of SN 2008S and N300OT over the same time period with \emph{HST}/WFC3 IR F110W and F160W imaging (GO-12331, 12450, 13613, and 14049).  We also use public WFC3 UVIS F438W, F606W, and F814W images of SN 2008S taken in Feb. 2014 (PI B. Sugerman, GO-13392) to supplement our optical limits.
% gethead WREG/*fits PROGID OBSRVR
% from: /data/poohbah/0/ckochanek/LBTmonitor/N6946/SST/ISISsmall_adams2
%   and other places...

\begin{figure}
  \ifpdflatex
    \includegraphics[width=8.6cm, angle=0]{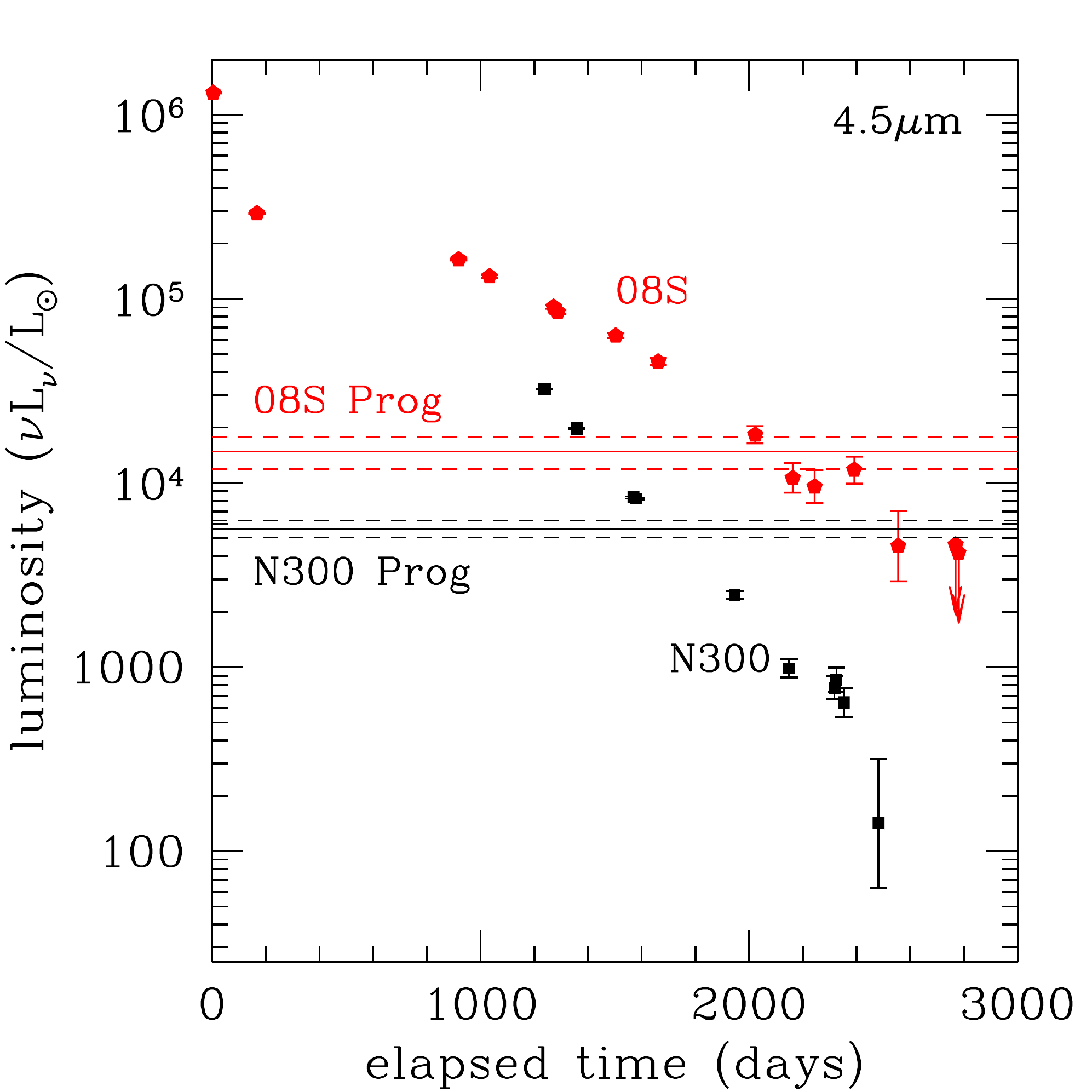}
  \else
    \includegraphics[width=8.6cm, angle=0]{fig1.ps}
  \fi
  \caption{The 4.5$~\mu\mathrm{m}$ light curves of SN 2008S (red pentagons) and N300OT (black squares).  The progenitor luminosities are given by the solid horizontal lines with the 1$\sigma$ uncertainties shown with dashed lines.  Both objects have faded below their progenitor luminosities at 4.5$~\mu\mathrm{m}$.  The late-time declines of $L_{4.5}$ are well-fit by an exponential for both sources.
\label{fig:ir_lc}}
\end{figure}
% see /data/poohbah/0/ckochanek/LBTmonitor/N6946/SST/ISISsmall_adams2/NOTES
%  /home/poseidon/sadams/scratch/sm plot_08s_n300_sys_poster_ad.sm
% ps2pdf -dEPSCrop lc_thick_oct2015.ps lc_thick_oct2015.pdf

We have been monitoring NGC 6946 (the host of SN 2008S) with the Large Binocular Camera \citep[LBC;][]{Giallongo08} on the LBT as part of a program searching for failed SNe \citep{Kochanek08,Gerke15}, with 30 epochs since 2008 in the U, B, V, and R bands.  We also use pre-eruption LBT B and V band images of NGC 6946 taken as part of a public program (PI Pasquali) in May 2007.
We monitored the optical evolution of N300OT with 6 epochs of R-band imaging taken with the Inamori-Magellan Areal Camera and Spectrograph \citep[IMACS;][]{Dressler11} on the Baade-Magellan 6.5-m telescope between 2009 and 2015.

Because we have image sequences with the transient varying, source location and identification is generally trivial.
Image subtraction was performed using {\sc isis} \citep{Alard98,Alard00} to measure variability in all filters for which we had multiple epochs.  Systematic uncertainties are estimated by generating {\sc isis} light curves for a grid of points within a certain separation (2.6--55" depending on the instrument) of the target, calculating the standard deviation of each light curve after 3$\sigma$ clipping, and then taking the average of these standard deviations after 3$\sigma$ clipping the ensemble as well.
% 400 pixels for LBT and Magellan
%       Magellan pixel scale is 0.2"/pix
%       LBT pixel scale is 0.225"/pix
%       so technically the systematics were estimated by a 45" grid for LBT and 40" grid for Magellan
% 200 pixel radius for HST
%       HST WFC3/IR pixel scale is ~0.128" = 25.6" grid
%       The HST grid is not centered on SN 2008S since it is close to the image edge...
% 90 pixel radius for SST
%       SST pixel scale is 0.6"/pix = 54"
The \emph{SST} light curves are given in Tables \ref{tab:SN2008Ssst} and \ref{tab:N300OTsst}.

\begin{figure}
  \ifpdflatex
    \includegraphics[width=8.6cm, angle=0]{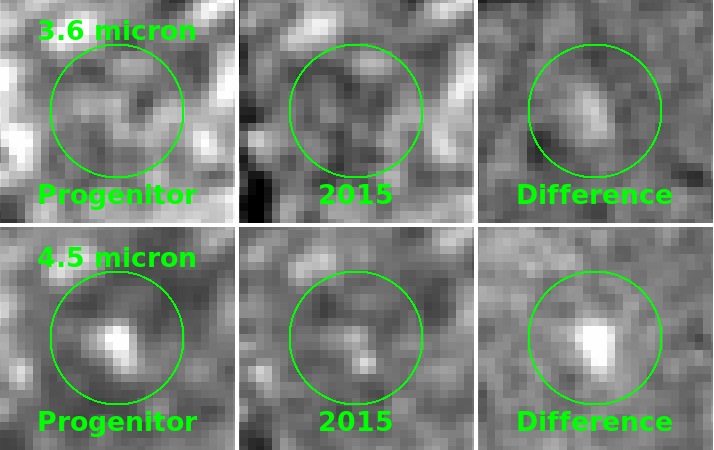}
    %\includegraphics[width=8.6cm, angle=0]{/home/poseidon/sadams/impostors/poster/08S/08s_ir_oct2015.png}
  %\else
  %  \includegraphics[width=8.6cm, angle=0]{/home/poseidon/sadams/impostors/poster/08S/08s_ir_oct2015.eps}
  \fi
  \caption{\emph{SST} images of the region surrounding SN 2008S.  The top row shows the 3.6$~\mu\mathrm{m}$ images and the bottom row shows the 4.5$~\mu\mathrm{m}$ images.  The left-hand panels are pre-eruption images, the center panels are the latest epochs, and the right-hand panels are the difference between the two, where flux decreases are white.  Each green circle is centered on the transient location and has a 5" radius.  The difference images show that SN 2008S is now fainter than its progenitor at both 3.6 and 4.5$~\mu\mathrm{m}$. \label{fig:08S_diff}}
\end{figure}
% bash 08s_diff.sh
% run from: /home/poseidon/sadams/impostors/poster/08S

\begin{figure}
  \ifpdflatex
    \includegraphics[width=8.6cm, angle=0]{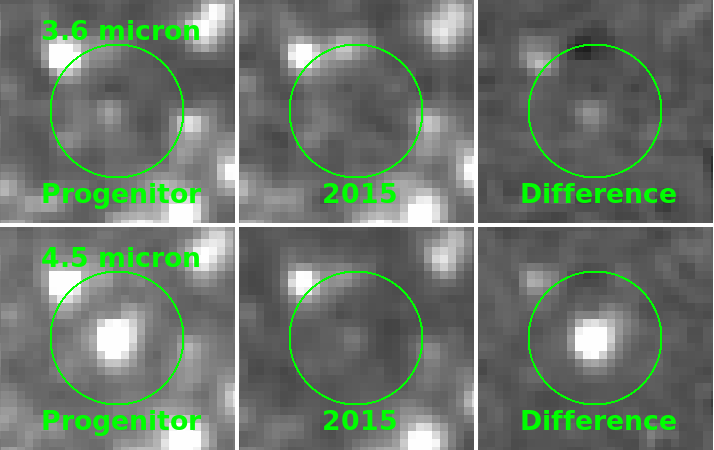}
  \else
    \includegraphics[width=8.6cm, angle=0]{fig3.eps}
  \fi
  \caption{\emph{SST} images of the region surrounding N300OT.  The top row shows the 3.6$~\mu\mathrm{m}$ images and the bottom row shows the 4.5$~\mu\mathrm{m}$ images.  The left-hand panels are pre-eruption images, the center panels are the latest epochs, and the right-hand panels are the difference between the two, where flux decreases are white.  Each green circle is centered on the transient location and has a 5" radius.  The difference images show that N300OT is now fainter than its progenitor at both 3.6 and 4.5$~\mu\mathrm{m}$. \label{fig:300ot_diff}}
\end{figure}
% bash n300_diff.sh
% run from: /home/poseidon/sadams/impostors/poster/08S

We also performed aperture photometry on the new \emph{SST} and \emph{HST} images and present the results in Tables \ref{tab:SN2008Sphotometry} and \ref{tab:NGC300photometry}.  For the \emph{SST} data we used a 2.4 arcsec radius aperture with a 2.4--4.8 arcsec radius sky annulus (except for the late-time SN 2008S data where we used a 1.2 arcsec radius aperture with a 1.2-2.4 arcsec radius sky aperture because the source had faded to a similar flux as a nearby source that would have contaminated the larger aperture) and the standard aperture corrections from the IRAC instrument handbook.  For the \emph{HST} data we used a 0.26 arcsec radius aperture with a 0.26--1.03 arcsec radius sky annulus and aperture corrections calculated using the \emph{HST} point spread function (PSF) models from {\sc tiny tim} \citep{Krist95,Krist11}\footnote{http://tinytim.stsci.edu/cgi-bin/tinytimweb.cgi}.  For comparison, the constraints on the progenitors of SN2008S and N300OT from previous works are given in Tables  \ref{tab:prog08s} and \ref{tab:n300prog}.

\begin{figure*}
  \ifpdflatex
    \includegraphics[width=0.9\textwidth]{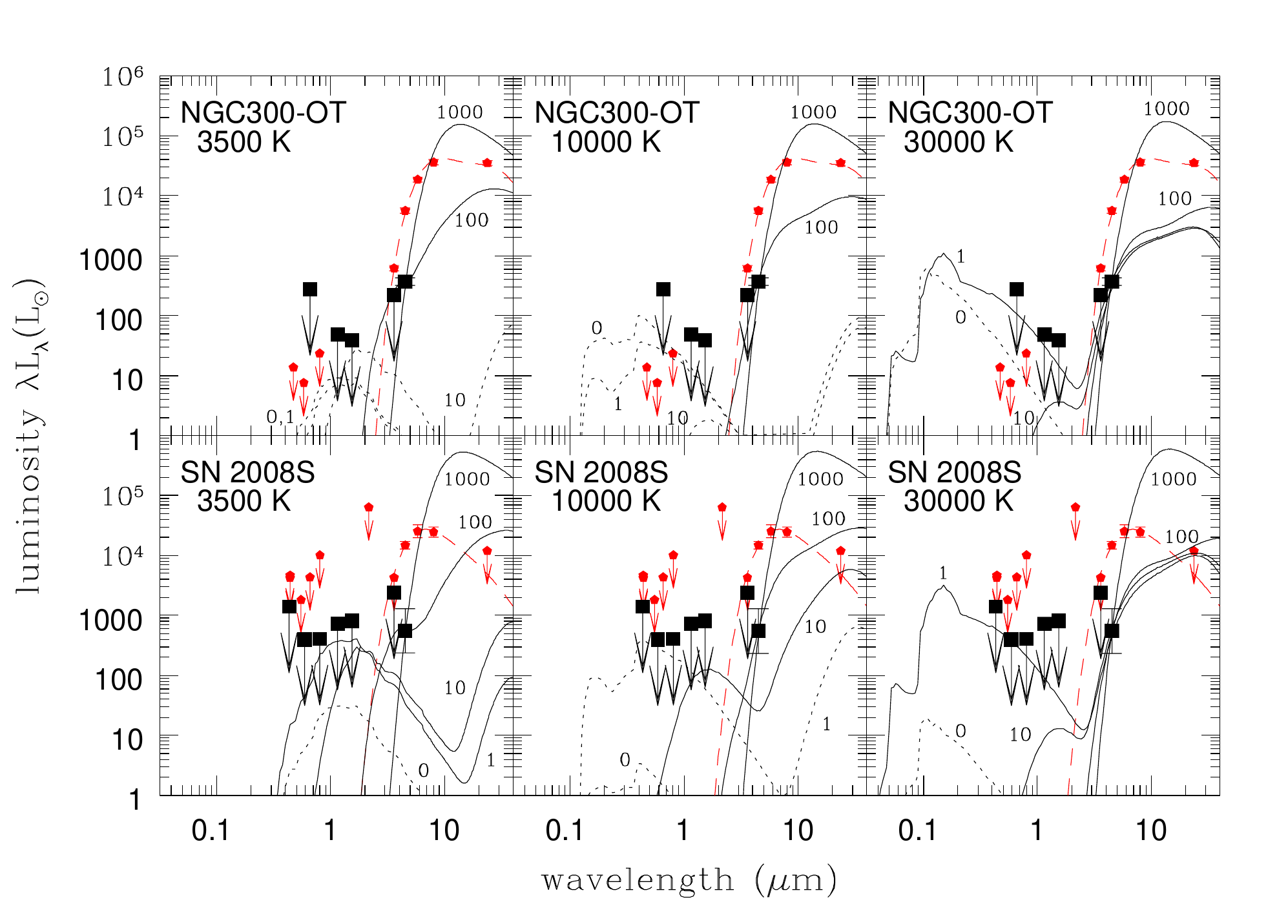}
  \else
    \includegraphics[width=0.9\textwidth]{fig4.ps}
  \fi
  \caption{Best-fit SEDs for possible surviving stars with $T_{*}=3500$, 10000, and 30000 K in the dusty shell scenario for $\tau_{V,\mathrm{tot}}=0$, 1, 10, 100, and 1000 are given by the labeled black lines, which are solid for N300OT and SN 2008S models consistent with the data and dotted if they are not.  The current photometric constraints are given by the large black squares.  For comparison the SEDs of the progenitors are displayed by the red dashed lines and small pentagons.  Many of the uncertainties for the detections are smaller than the sizes of the points.
Since the only late-time detections are at $4.5~\mu\mathrm{m}$ and there are no constraints at longer wavelengths there are some degeneracies in the solutions, with increasingly luminous survivors allowed for high optical depths and cooler dust photospheres.  The progenitor models were fixed to $T_{*} = 3500$ K and $T_{\mathrm{dust}}=1500$ K and the dusty shell models were fixed to $R_{\mathrm{out}}/R_{\mathrm{in}}=2$.  Consistency with the data is defined here by $\chi^{2}<16.8$ (26.2) for N300OT (SN 2008S), where the probability of exceeding the $\chi^{2}$ value is $1\%$ for 6 (12) degrees of freedom.  Surviving stars with luminosities similar to the progenitors require $\tau_{V,\mathrm{tot}}\gtrsim100$.
%The progenitor model parameters are $\tau_{V,\mathrm{tot}}=650$, $R_{\mathrm{out}}/R_{\mathrm{in}}=22$, and $L_{*} = 10^{4.86}~L_{\odot}$ with $\chi^2=8.3$ for N300OT and
%$\tau_{V,\mathrm{tot}}=290$, $R_{\mathrm{out}}/R_{\mathrm{in}}=5.8$, and $L_{*} = 10^{4.52}~L_{\odot}$ with $\chi^2=1.6$ for SN 2008S.
\label{fig:seds}}
\end{figure*}

Figures \ref{fig:ir_lc}, \ref{fig:08S_diff}, \ref{fig:300ot_diff}, and \ref{fig:seds} provide illustrations of the fundamental observation that both transients are now fainter than their progenitors at every band where it is presently feasible to make the comparison.  Fig. \ref{fig:ir_lc} shows this using the 4.5$~\mu\mathrm{m}$ light curves and Figures \ref{fig:08S_diff} and \ref{fig:300ot_diff} illustrate this visually using pre-transient and present-day \emph{SST} images, along with their differences.
The current SED constraints are compared to those of the progenitors in Fig. \ref{fig:seds}.  As before the transients, there are now only upper limits on the optical and near-IR fluxes of the two sources.

We can also use image subtraction to set stringent limits on the late-time variability of SN 2008S and N300OT.
Tables \ref{tab:SN2008Svariability} and \ref{tab:NGC300variability} list the constraints found when only using data taken after the transients had completely faded ($\gtrsim2$ and $\gtrsim4$ years post-peak for the optical and near-IR, respectively).

\subsubsection{X-ray}
\label{sec:xray}
We observed SN 2008S with ACIS-S \citep{Garmire03} onboard the \emph{Chandra X-ray Observatory} \citep{Weisskopf02} on 2012--5--21 with an exposure time of 20.4~ks.
The data were reduced following the standard procedures.
We reprocessed all the data using the CIAO 4.7 software tools, and the events are filtered using the standard ASCA grades of 0, 2, 3, 4, and 6.
We did not detect the X-ray counterpart of SN~2008S, and set a 90\% confidence upper limit on the absorbed (i.e., observed) flux of $1.4\times10^{-15}~\mathrm{erg}\>\mathrm{s}^{-1}\>\mathrm{cm}^{-2}$ in the full 0.5--7~keV band, and $3.8\times10^{-16}$, $4.7\times10^{-16}$, and $2.4\times10^{-15}~\mathrm{erg}\>\mathrm{s}^{-1}\>\mathrm{cm}^{-2}$ in the 0.5--1.2, 1.2--2, and 2--7~keV bands.
To correct for absorption we adopt log $N_{H} = 21.6$ for the Galactic absorption to X-ray sources in NGC 6946 \citep{Holt03}.  This an order of magnitude higher than the value reported by {\sc colden}\footnote{http://cxc.harvard.edu/toolkit/colden.jsp}, the \emph{Chandra} Galactic neutral hydrogen density calculator.  
Additionally, we estimate the hydrogen column density of the CSM exterior to the expanding shock assuming the wind density parameter of the progenitor found in \S\ref{sec:dusty} as
\begin{eqnarray}
N_{\mathrm{H}} \simeq 10^{22.69} \left(\frac{\dot{M}/v_{\mathrm{w}}}{2.4\times10^{-5}~M_{\odot}\>\mathrm{yr}^{-1}/\mathrm{km}\>\mathrm{s}^{-1}}\right) \nonumber \\ \times \left(\frac{1100~\mathrm{km}\>\mathrm{s}^{-1}}{v_{\mathrm{ej}}}\right) \left(\frac{4.3~\mathrm{yr}}{t_{\mathrm{elap}}}\right)~\mathrm{cm}^{-2} ,
\end{eqnarray}
where $\dot{M}$ is the progenitor mass loss rate, $v_{\mathrm{w}}$ is the velocity of the progenitor wind, $v_{\mathrm{ej}}$ is the ejecta velocity, and $t_{\mathrm{elap}}$ is the time elapsed since the start of the transient at the epoch of X-ray observation.
The upper limit on the IR luminosity that could come from absorbed X-rays given our X-ray non-detection is
\begin{equation}
L_{IR} < \left(\frac{4\pi d^{2} F_{\mathrm{obs,Chandra}}}{f_{\mathrm{Chandra}}}\right) \left(\frac{1-f_{\mathrm{trans,CSM}}}{f_{\mathrm{trans,CSM}} f_{\mathrm{trans,gal}}}\right) ,
\end{equation}
where $F_{\mathrm{obs,Chandra}}$ is the observed flux in the \emph{Chandra} bandpass, $d$ is the distance to NGC 6946, $f_{\mathrm{Chandra}}$ is the fraction of the X-ray luminosity emitted in the \emph{Chandra} bandpass, $f_{\mathrm{trans,CSM}}$ is the fraction of the X-ray luminosity transmitted through the column density of the CSM exterior to the shock, and $f_{\mathrm{trans,gal}}$ is the fraction of the remaining X-ray luminosity transmitted through the column density of the Galaxy.
The characteristic X-ray energy of a shock moving through a wind at velocity $v_{\mathrm{s}}$ is
\begin{equation}
E_{\mathrm{s}} = \frac{3\mu}{16}m_{\mathrm{p}}v_{\mathrm{s}}^{2} \simeq 1.46 \left(\frac{v_{\mathrm{s}}}{1100~\mathrm{km}\>\mathrm{s}^{-1}}\right)^{2}~\mathrm{keV}
\end{equation}
for a mean molecular weight of $\mu=0.6$.
Using the {\sc pimms} calculator\footnote{http://cxc.harvard.edu/toolkit/pimms.jsp} for \emph{Chandra} with a 1.46 keV shock and a thermal Bremsstrahlung model to estimate $f_{\mathrm{Chandra}}$, $f_{\mathrm{trans,CSM}}$, and $f_{\mathrm{trans,gal}}$, we find
\begin{eqnarray}
L_{IR} < 2.5\times 10^{4} \left(\frac{d}{5.6~\mathrm{Mpc}}\right)^{2} \left(\frac{F_{\mathrm{obs,Chandra}}}{1.4\times10^{-15}~\mathrm{erg}\>\mathrm{s}^{-1}\>\mathrm{cm}^{-2}}\right) \nonumber \\ \times \left(\frac{0.62}{f_{\mathrm{Chandra}}}\right) \left(\frac{f_{\mathrm{trans,CSM}}^{-1}-1}{6.0}\right) \left(\frac{0.62}{f_{\mathrm{trans,gal}}}\right)~L_{\odot} .
\label{eqn:Xraylim}
\end{eqnarray}

% first find best Lstar and tdust for each tau & Tstar combo using: python ../scripts/findbest.py
%   from /home/poseidon/sadams/impostors/dusty
% then create sed from /home/poseidon/sadams/impostors/dusty/SN2008S/seds
%  and /home/poseidon/sadams/impostors/dusty/ngc300ot/seds
% directories with dusty_wrapper.py
% then create plot using plot2x6.sm
%   from /home/poseidon/sadams/impostors/dusty/SN2008S/seds
% then manually trim the bounding box with vim
%       from: 
%               %%BoundingBox: 18 144 592 718
%       to:
%               %%BoundingBox: 18 148 592 560
% ps2pdf -dEPSCrop seds.ps seds.pdf

\subsection{Dust Models \& Scalings}
\label{sec:dustmodels}

It is necessary to consider plausible models of dust obscuration when interpreting the significance of the observed decline in mid-IR flux of the targets and the non-detections at optical and near-IR wavelengths.  The evolution of the SEDs during the transients indicate that dust re-formed in the pre-existing winds \citep{Kochanek11b}, but this, at most, can only return the objects to their original level of obscuration.  
If the progenitors survived without a decrease in bolometric luminosity, the observed decrease in mid-IR flux requires an additional source of obscuration.

We focus on two possibilities for increased obscuration: 
dust formed in a shell ejected at the time of the explosion expanding into the pre-existing wind or dust formed in a thicker wind that began following the transient.
These scenarios and the dust scalings we present below are similar to the analysis we described in \citet{Kochanek12} and \citet{Adams15}.

In the shell model, dust would form once the ejecta has expanded far enough from the star for it to be cool enough for dust to condense at the dust formation temperature, $T_{f}$.  
The continuing expansion of the shell drives down its density so that most of the dust is formed soon after the shell reaches the radius, $R_{f}$, at which dust formation can occur.  
At late-times, after dust formation is complete, this geometric dilution (assuming constant velocity) translates into an optical depth that decreases as 
\begin{equation}
\tau(t) = \frac{M_{\mathrm{ej}}\kappa}{4\pi v_{\mathrm{e}}^{2} t^{2}} , 
\label{eqn:taut}
\end{equation}
where $M_{\mathrm{ej}}$ is the ejected mass, $v_{\mathrm{e}}$ is the radial velocity of the ejected shell and $\kappa$ is the opacity.  Any surviving star that is obscured by a dusty, ejected shell should appear to re-brighten as the optical depth decreases.  Deviations from this simple scaling for spherical expansion can generally only accelerate the drop in the optical depth \citep[see][]{Kochanek12}.  Since the current observed luminosity in a given filter, $\mathrm{f}$, is $L_{\mathrm{f,obs}}=L_{\mathrm{f},*} e^{-\tau_{\mathrm{f,eff}}}$, where $L_{\mathrm{f},*}$ is the stellar luminosity in band $\mathrm{f}$, and $\tau_{\mathrm{f,eff}}$ is the current effective optical depth in the given filter, 
the limit on the observed rate of change in the flux in a given band, $dL_{\mathrm{f,obs}}/dt$ constrains the maximum luminosity, $L_{\mathrm{f},*}$, of a surviving star within the expanding shell to
\begin{equation}
L_{\mathrm{f},*}< \frac{1}{2}\frac{t}{\tau_{\mathrm{f,eff}}}\left(\frac{dL_{\mathrm{f,obs}}}{dt}\right) e^{\tau_{\mathrm{f,eff}}} .
\label{eqn:lumlimit}
\end{equation}
In this model, the mass of the ejected shell required for a given, current, total optical depth, $\tau_{\mathrm{f,tot}}$, is
\begin{equation}
M_{\mathrm{ej}} = \frac{ 4 \pi v_{\mathrm{e}}^{2} t^{2} \tau_{\mathrm{f,tot}}(t)}{\kappa_{\mathrm{f}}} .
\label{eqn:shellmass}
\end{equation}
The total and effective optical depths are related by $\tau_{\mathrm{eff}} = \left[ \tau_{\mathrm{abs}} (\tau_{\mathrm{abs}} + \tau_{\mathrm{sca}}) \right]^{1/2}$, with $\tau_{\mathrm{abs}}$ and $\tau_{\mathrm{sca}}$ being the absorption and scattering optical depths, respectively.  This relation can also be expressed in terms of the scattering albedo, $w$, by $\tau_{\mathrm{eff}} = (1-w)^{1/2} \tau_{\mathrm{tot}}$.

\begin{figure}
  \ifpdflatex
     \includegraphics[width=8.6cm, angle=0]{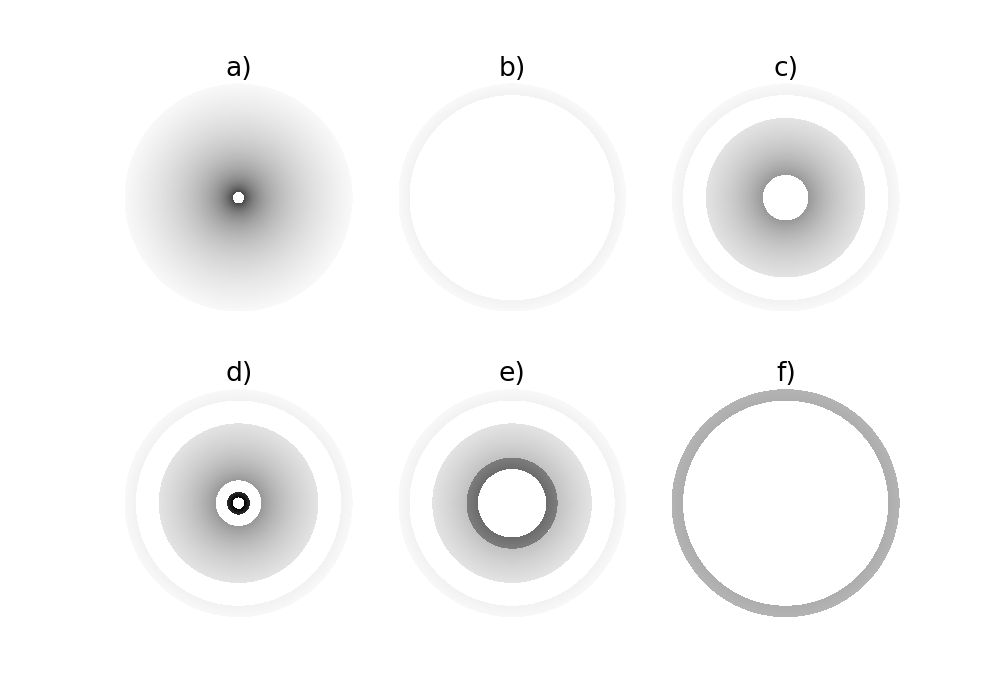}
  \else
    \includegraphics[width=8.6cm, angle=0]{fig5.ps}
  \fi
  \caption{Schematic of the possible evolution of dust around SN 2008S-like transients:  a) Pre-transient---the progenitor is obscured by a thick, dusty wind,  b) the transient destroys most of the dust from the inside out,  c) dust starts to re-form in the wind and obscures the transient,  d) dust forms in the shell ejected during the transient after it reaches $R_{\mathrm{f}}$, and  e \& f)  the ejected shell expands and destroys the dust formed in the pre-existing wind as the shock front passes through. \label{fig:schematic}}
\end{figure}

As the SN ejecta expands into the dusty CSM the pre-existing dust is likely destroyed by the passage of the shock front \citep{Draine79,Slavin15}.  A schematic illustration of the evolution of the obscuration is given in Fig. \ref{fig:schematic}.
If this is the case, the optical depth would evolve more quickly than $t^{-2}$ and Equation \ref{eqn:lumlimit} would still be a valid upper limit on the luminosity of a surviving star.

Alternatively, the optical depth may avoid the $\tau(t) \propto t^{-2}$ evolution of the shell case if the obscuration is dominated by dust in a steady state wind.  If we again assume that all of the dust forms at the dust formation radius, $R_{\mathrm{f}}$, the rate of mass loss for a steady wind of a given optical depth and extending to infinity is
\begin{equation}
\dot{M} = \frac{4\pi v_{\mathrm{w}} R_{\mathrm{f}} \tau_{\mathrm{f,tot}}}{\kappa_{\mathrm{f}}} ,
\label{eqn:windmass}
\end{equation}
where $R_{\mathrm{f}} \sim L^{1/2}/T_{\mathrm{f}}^{2}$ and $T_{\mathrm{f}}\sim1500~\mathrm{K}$.

We will use these relations in \S\ref{sec:results} to help determine whether there could be surviving stars to SN 2008S and N300OT obscured by an expanding shell or steady-state wind.  

% generated by: python schematic.py
%  run from: /home/poseidon/sadams/impostors/scripts
% then used: convert schematic.png schematic.ps
% (the program doesn't like saving .ps when using a high resolution grid)

\subsection{DUSTY}
\label{sec:dusty}

Following the methods of \citet{Adams15}, we model the SEDs of SN 2008S and N300OT with the dusty radiative transfer code {\sc dusty} \citep{Ivezic97,Ivezic99,Elitzur01}, with stellar atmospheric models from \cite{Castelli04} for stars of solar composition but with various temperatures.
We present possible model SEDs found by using a Markov Chain Monte Carlo wrapper around {\sc dusty} in Fig. \ref{fig:seds}.  
The shell models assume expansion velocities of $v_{\mathrm{e}} = 560$ and $1100~\mathrm{km\>s^{-1}}$ \citep[$\pm$ a factor of two;][]{Smith11} for NGC 300-OT and SN 2008S respectively, with the shell ejected at the beginning of the transients.  
We incorporate the variability constraints (listed in Tables \ref{tab:NGC300variability} and \ref{tab:SN2008Svariability}) into the shell model by adding $\chi^{2}$ contributions for each constrained filter, f, found by
\begin{equation}
\chi^2_{\mathrm{f}} = \left( \frac{dL_{\mathrm{f,obs}}/dt - dL_{\mathrm{f,mod}}/dt}{\sigma_{dL_{\mathrm{f,obs}}/dt}} \right)^2 ,
\end{equation}
where the model variability, $dL_{\mathrm{f,mod}}/dt$, is
\begin{equation}
\frac{dL_{\mathrm{f,mod}}}{dt} = \frac{2 L_{\mathrm{f}} \tau_{\mathrm{f,eff}}}{t_{\mathrm{elap}}} .
\end{equation}
Unless noted otherwise, we assume a dust shell thickness of $R_{\mathrm{out}}/R_{\mathrm{in}} = 2$.
We assume graphitic dust \citep[based on][]{Prieto09} and an MRN grain size distribution (which has an optical albedo of $w_V\simeq 0.47$, corresponding to $\tau_{V,\mathrm{eff}}\simeq 0.73 \tau_{V,\mathrm{tot}}$).
For the wind case, the inner edge of the dust distribution is set by $R_{\mathrm{f}}$.  
Since emission is only detected at 4.5$~\mu\mathrm{m}$ and there are no limits at longer wavelengths, there are many possible ways to model the SED.  In \S\ref{sec:results} we present MCMC results for three representative stellar temperatures ($T_{*}=3500$, 10000, and 30000 K).  For completeness we also provide results for grids of models with different stellar temperatures and optical depths in Appendix \ref{app:one}.

We also generate new fits for the progenitors of SN 2008S and N300OT as super-AGB stars obscured in constant velocity winds of graphitic dust.
Fixing $T_{*} = 3500$ K and $T_{\mathrm{dust}}=1500$ K,
gives $\mathrm{log}~L_{*}/L_{\odot} = 4.54 \pm 0.07$, $\tau_{V,\mathrm{tot}}= 290^{+250}_{-70}$,
log $R_{\mathrm{f}}/\mathrm{cm} = 14.62^{+0.08}_{-0.05}$,
log $R_{\mathrm{phot},4.5}/\mathrm{cm} =15.1\pm0.1$, and
$\mathrm{log}(\dot{M}/v_{\mathrm{w}}) = -4.6_{-0.2}^{+0.4}$ for SN 2008S and $\mathrm{log}~L_{*}/L_{\odot} = 4.88 \pm 0.02$,
$\tau_{V,\mathrm{tot}}= 600^{+30}_{-20}$,
log $R_{\mathrm{f}}/\mathrm{cm} = 14.84\pm0.02$,
log $R_{\mathrm{phot},4.5}/\mathrm{cm} =15.74\pm0.03$, and
$\mathrm{log}(\dot{M}/v_{\mathrm{w}}) = -4.09_{-0.04}^{+0.03}$ for N300OT,
where $R_{\mathrm{phot},4.5}$ is the $4.5\>\mu\mathrm{m}$ photospheric radius, $\dot{M}$ is in $M_{\odot}\>\mathrm{yr}^{-1}$ and $v_{\mathrm{w}}$ is in km s$^{-1}$. 
  These values of $\dot{M}_{\odot}\>\mathrm{yr}^{-1}$ are shifted relative to \citet{Kochanek11b} because of the use of constant velocity winds rather than than {\sc dusty}'s self-consistent wind acceleration models.
% python ~/impostors/dusty/ngc300ot/evolution/paramevol.py progenitor_tstar3500_td1500_noext/results.dat tau tau --burn 150
% python ~/impostors/dusty/ngc300ot/evolution/paramevol.py progenitor_tstar3500_td1500_noext/results.dat lstar lstar --burn 150

\section{Surviving Stars Obscured by Dust?}
\label{sec:results}

\subsection{Dusty Shell}
\label{sec:shellresults}

Figure \ref{fig:seds} illustrates the constraints on possible surviving stars.  We discuss three representative possibilities for the stellar temperature: an AGB star with $T_{*}=3500$ K, a hotter star with $T_{*}=10^{4}$ K \citep[e.g. a star on a ``blue loop" as suggested by][]{Humphreys11}, and a still hotter star with $T_{*}=3\times10^{4}$ K.  Since the progenitors were likely cool, super-AGB stars \citep{Thompson09}, we first consider the constraints on such stars (see the left-hand column of Fig. \ref{fig:seds}).  
The SED from an unobscured 3500 K star peaks around $1~\mu\mathrm{m}$ and is strongly constrained by the very deep optical and near-IR limits. 
If the $4.5~\mu\mathrm{m}$ flux is reprocessed emission from a surviving star, the deep constraints at shorter wavelengths require that any stellar source be obscured by $\tau_{V,\mathrm{tot}}>100$ for N300OT and $>10$ for SN 2008S.
However, stars at these lower optical depths also have to be significantly less luminous than the progenitors.

\begin{figure*}
  \ifpdflatex
    \includegraphics[width=0.95\textwidth]{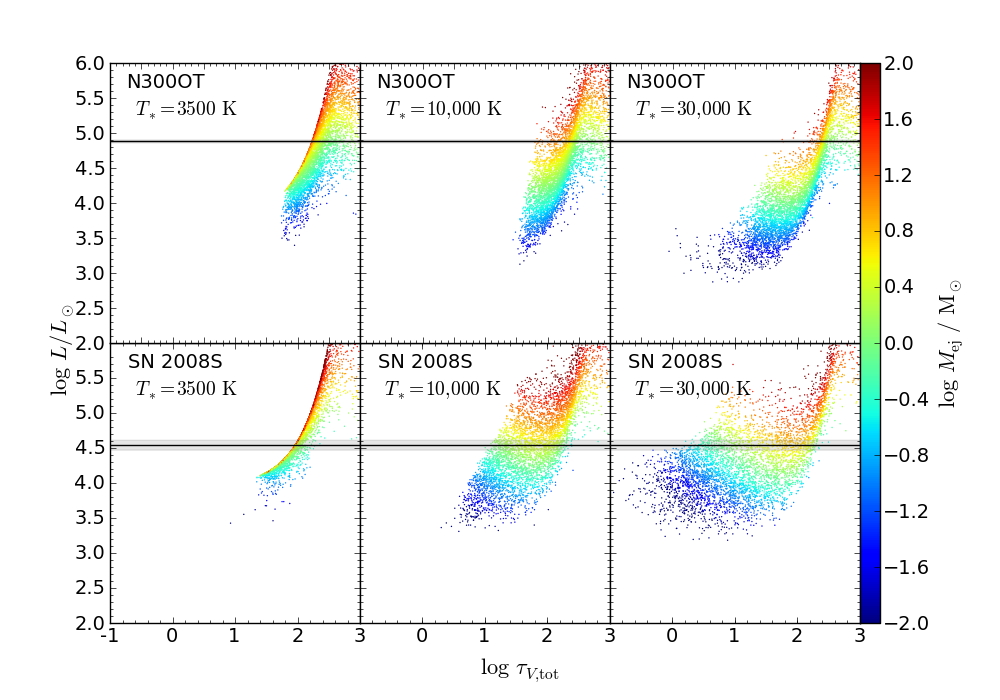}
  \fi
  \caption{MCMC results for the luminosity and dust optical depth of a surviving star obscured by an expanding shell ejected at the time of the optical transients for N300OT (top panels) and SN 2008S (bottom panels).  For comparison the progenitor luminosity and $1\sigma$ uncertainties are displayed by the horizontal black line and surrounding gray shading.  The MCMC points are color-coded by the ejected mass implied by Eqn. \ref{eqn:shellmass} assuming $\kappa_{V}=100~\mathrm{cm}^{2}\>\mathrm{g}^{-1}$.}
\label{fig:fixedTs}
\end{figure*}
% generated by: python fixedTs.py
%   run from: /home/poseidon/sadams/impostors/dusty
%       SN 2008S T=3500: 84.8 < tau < 110.9
%       SN 2008S T=3500: 608.9 < v < 2666.7
%       SN 2008S T=3500: 1.5 < Mej < 24.2
%       SN 2008S T=10000: 18.7 < tau < 160.1
%       SN 2008S T=10000: 529.9 < v < 1474.3
%       SN 2008S T=10000: 1.2 < Mej < 3.2
%       SN 2008S T=30000: 2.9 < tau < 150.2
%       SN 2008S T=30000: 755.5 < v < 2138.9
%       SN 2008S T=30000: 0.4 < Mej < 5.6
%       N300OT T=3500: 174.2 < tau < 486.0
%       N300OT T=3500: 274.4 < v < 1610.8
%       N300OT T=3500: 1.0 < Mej < 13.0
%       N300OT T=10000: 124.2 < tau < 518.5
%       N300OT T=10000: 270.6 < v < 1909.1
%       N300OT T=10000: 1.1 < Mej < 12.6
%       N300OT T=30000: 166.8 < tau < 590.4
%       N300OT T=30000: 263.9 < v < 1917.3
%       N300OT T=30000: 1.1 < Mej < 17.7

For $T_{*}=10000$ (middle column of Fig. \ref{fig:seds}) the required optical depths are only slightly reduced.  
The optical limits strongly restrict luminous solutions at low optical depths and the $\tau_{V,\mathrm{tot}}=0$, 1, and 10 models are unable to fit the $4.5~\mu\mathrm{m}$ flux without violating the optical and near-IR limits.
Only if the stars are made still hotter, as illustrated by the $T_{*}=30000$ K model (right-hand column of Fig. \ref{fig:seds}), do the optical limits become less constraining.  However, large optical depths are still required for the $4.5~\mu\mathrm{m}$ flux to be from reprocessed stellar emission.  

If viewed only as limits, the late-time photometric data do not, by themselves, rule out the possibility of very hot ($\gtrsim30,000~\mathrm{K}$ and $\gtrsim45,000~\mathrm{K}$ for SN 2008S and N300OT respectively), unobscured stars with the luminosity of the progenitors.  However, the evolution of the transient SEDs place lower limits on the current optical depths.  \citet{Kochanek11b} found that the optical depths had evolved to $\tau_{V}\sim100$ by 1000 days.  Even if no additional dust formed and we include the $\tau \propto t^{-2}$ evolution from geometric expansion, the current optical depths would be $15<\tau_{V}<20$ while an optical depth of just $\tau_{V} \sim 1$ would reprocess enough UV flux from a hot star into longer wavelengths that the photometric limits would not allow a hot surviving star with the progenitor luminosity.

We now consider the mass loss needed to produce the high optical depths required by the SED constraints for surviving stars.  Figure \ref{fig:fixedTs} shows the MCMC results for surviving stars with different temperatures obscured by a dusty shell.  We estimate the mass loss implied by these models using Equation \ref{eqn:shellmass}.  Higher optical depths can hide  more luminous stars.  At a given stellar luminosity, the spread in the optical depth comes from the factor of two uncertainty used for the shell expansion velocity.  A faster expansion velocity corresponds to a larger, cooler dust photosphere that emits more radiation at wavelengths longer than our $4.5~\mu\mathrm{m}$ constraint and thus does not need as high of an optical depth.  However, the required ejected mass is generally larger for these solutions because the ejected mass is also proportional to the area of the shell.  The $T_{*}=3500$ K models require the ejected mass of the shell to be $>1.0$ and $>1.5~M_{\odot}$ (at the $90\%$ confidence level) for N300OT and SN 2008S, respectively.  Allowing for the surviving stars to have become hotter does not significantly change the required ejected mass for N300OT and only decreases it for SN 2008S to 1.2 and $0.4~M_{\odot}$ for the 10000 and 30000 K cases, respectively.

The expanding shells would sweep up the pre-existing CSM (as shown schematically in Fig. \ref{fig:schematic}).  The passage of the shock would destroy dust that has re-formed in the pre-existing wind \citep{Draine79,Slavin15}, but this swept up material could potentially form dust yet again in the expanding shell.  
If the shocks have been freely expanding up until our latest observations the swept up mass would be
\begin{eqnarray}
M_{\mathrm{sw}}= 0.3 \left(\frac{\dot{M}/v_{\mathrm{w}}}{8\times10^{-5}~M_{\odot}\>\mathrm{yr}^{-1}\>/\>\mathrm{km}\>\mathrm{s}^{-1}} \right) \nonumber \\ \left(\frac{v_{\mathrm{e}}}{560~\mathrm{km}\>\mathrm{s}^{-1}}\right) \left(\frac{t_{\mathrm{elap}}}{6.8~\mathrm{yr}}\right)~M_{\odot}
\end{eqnarray}
for N300OT and $\sim0.2~M_{\odot}$ for SN 2008S.  With dust re-forming in the swept up wind, the $M_{\mathrm{ej}}$ inferred from Equation \ref{eqn:shellmass} needed to hide the surviving stars could be reduced by up to $M_{\mathrm{sw}}$.  Even after accounting for this, $M_{\mathrm{ej}}>0.7$ and $>1.2~M_{\odot}$ are needed to hide cool (3500 K) surviving stars for N300OT and SN 2008S, respectively. 

We can compare the required mass ejection with estimates based on the transient light curves.  \citet{Kochanek12} estimate ejected masses of 0.07 and $0.12~M_{\odot}$ by equating the diffusion time to the timescale for the luminosity to decrease by 1.5 magnitudes and ejected mass of 0.24 and $0.43~M_{\odot}$ from the photon ``tiring limit" for radiatively driven mass loss for N300OT and SN 2008S, respectively.  If the ratio of radiated to kinetic energy is similar to that of  $\eta$ Car's Great Eruption, the mass loss in the eruptions would be 0.1--1 $M_{\odot}$ for N300OT \citep{Humphreys11} and 0.05--0.2 for SN 2008S \citep{Smith09}.
The amount of mass ejected in the transient needed to hide a survivor with a dusty shell is in significant tension with most of these estimates.  Of course this is also an issue for interpreting the transients as SNe.

\subsection{Dusty Wind}
\label{sec:dustywind}

Given that the mid-IR fluxes and limits are lower than those of the progenitors, the stars have not simply re-enshrouded themselves with the wind emitted prior to the transients.  To answer whether the transients could have signaled a transition to a still higher-mass loss state that is obscuring the stars in a thicker wind, we first must establish that enough time has elapsed for this new wind to pass the dust formation radius.  For the luminosities of the progenitors, the dust formation radii are (initially) $10^{14.2-14.9}$ cm (for $3500~\mathrm{K}<T_{*}<39,000~\mathrm{K}$ and a dust condensation temperature of 1500 K)\footnote{The dust formation radius also depends on $\tau$ once the wind is optically thick because a forming dust grain is also heated from radiation reprocessed by dust beyond the dust formation radius.  By $\tau_{V}\sim1000$, $R_{\mathrm{f}}$ has increased to $10^{14.8-15.0}$ for the same parameters.}, which would require wind velocities of $10-40~\mathrm{km\>s^{-1}}$ in order for dust to have begun to form in a new wind.  The low end of these velocities is comparable to the winds of $15 \pm 4$ and $12 \pm 3~\mathrm{km\>s}$ inferred from {\sc dusty}'s self-consistent dust-driven wind models for SN 2008S and N300OT, respectively \citep{Kochanek11b}.  So it is plausible that obscuration from a denser wind has begun to develop 

\begin{figure*}
  \ifpdflatex
    \includegraphics[width=0.95\textwidth]{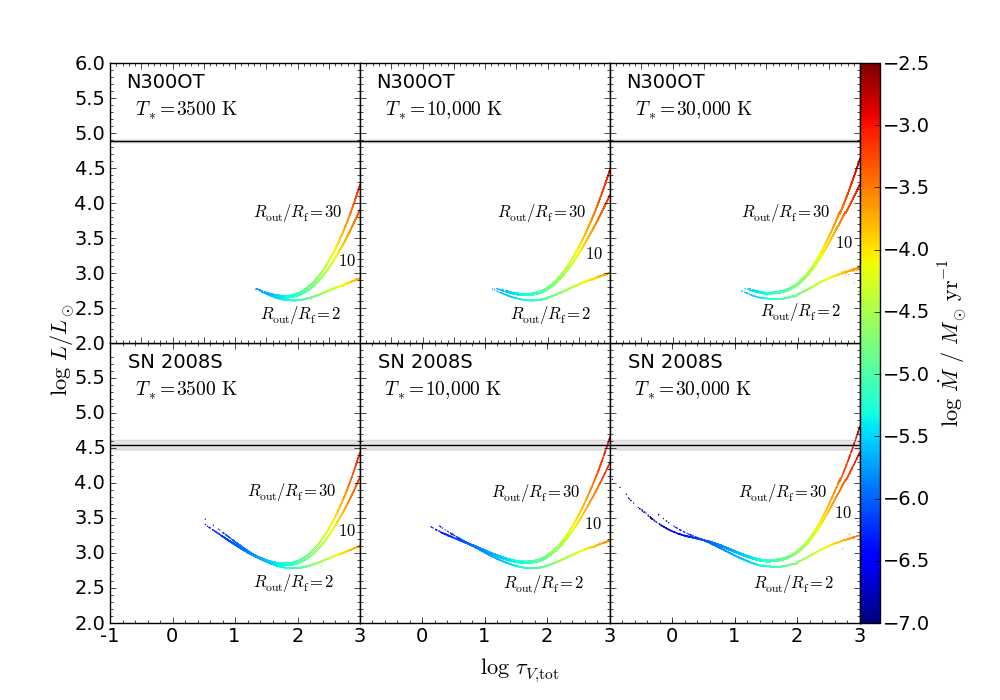}
    %\includegraphics[width=0.95\textwidth]{/home/poseidon/sadams/impostors/dusty/fixedTs_wind3ts.png}
  %\else
  %  \includegraphics[width=8.6cm, angle=0]{/home/poseidon/sadams/impostors/dusty/ngc300ot/evolution/rphot.eps}
  \fi
  \caption{MCMC results for the luminosity and dust optical depth of a surviving star obscured by a steady-state wind with $R_{\mathrm{out}}/R_{\mathrm{f}}=2$, 10 and 30 for N300OT (top row) and SN 2008S (bottom row).  Increasing $R_{\mathrm{out}}/R_{\mathrm{f}}$ further has little effect.  For comparison the progenitor luminosity and $1\sigma$ uncertainties are displayed by the horizontal black line and surrounding gray shading.  The MCMC points are color-coded by the mass loss implied by Eqn. \ref{eqn:windmass} assuming $\kappa_{V}=100~\mathrm{cm}^{2}\>\mathrm{g}^{-1}$ and a wind velocity of $10~\mathrm{km}\>\mathrm{s}^{-1}$.  A wind with this velocity starting at the onset of the optical transient has only had time to reach $\sim2R_{\mathrm{f}}$.  A thicker wind requires a proportionally higher $v_{\mathrm{w}}$ and $\dot{M}$ (unless dust has re-formed in pre-existing wind that was not swept up by a shock).  While $10~\mathrm{km}\>\mathrm{s}^{-1}$ is an appropriate wind velocity for a 3500 K star, a 10000 K or 30000 K star would be expected to have a wind velocity of hundreds or thousands of $\mathrm{km}\>\mathrm{s}^{-1}$ \citep[e.g.,][]{Kudritzki00} and the implied $\dot{M}$ should be scaled accordingly by the reader.  In all cases, luminous surviving stars require $\tau_{V,\mathrm{tot}}\gtrsim1000$ and very high mass loss rates.}
\label{fig:fixedTs_wind}
\end{figure*}
% generated by: python fixedTs_wind3ts.py
%   run from: /home/poseidon/sadams/impostors/dusty

Figure \ref{fig:fixedTs_wind} shows the MCMC results for surviving stars with different temperatures obscured by a steady-state wind, although we note that dust formation is unlikely for the two hotter models \citep[see][]{Kochanek11}.  First we consider the possibility that the surviving stars are cool, like the progenitors.  With $\tau_{V,\mathrm{tot}}$ up to 1000,
a wind with $R_{\mathrm{out}}/R_{\mathrm{f}}=2$ can only obscure surviving stars that are much fainter than the progenitors.  The main problem is that in order to radiate luminosities similar to the progenitor at wavelengths longer than our $4.5~\mu\mathrm{m}$ constraint, the radius of the $4.5~\mu\mathrm{m}$ photosphere, $R_{4.5}$, must be much larger than $2R_{\mathrm{f}}$.  
For a given optical depth, a wind that has extended to a larger radius is able to hide a more luminous star (compare the $R_{\mathrm{out}}/R_{\mathrm{f}}=30$, 10, and 2 results in Fig. \ref{fig:fixedTs_wind}) because the larger radius results in a cooler dust photosphere that is less constrained by our mid-IR limits.  However, the photometric constraints require the mid-IR photosphere of a luminous surviving star to be much larger than the radius that a post-transient, dust-driven wind could have reached.

\begin{figure}
  \ifpdflatex
    \includegraphics[width=8.6cm, angle=0]{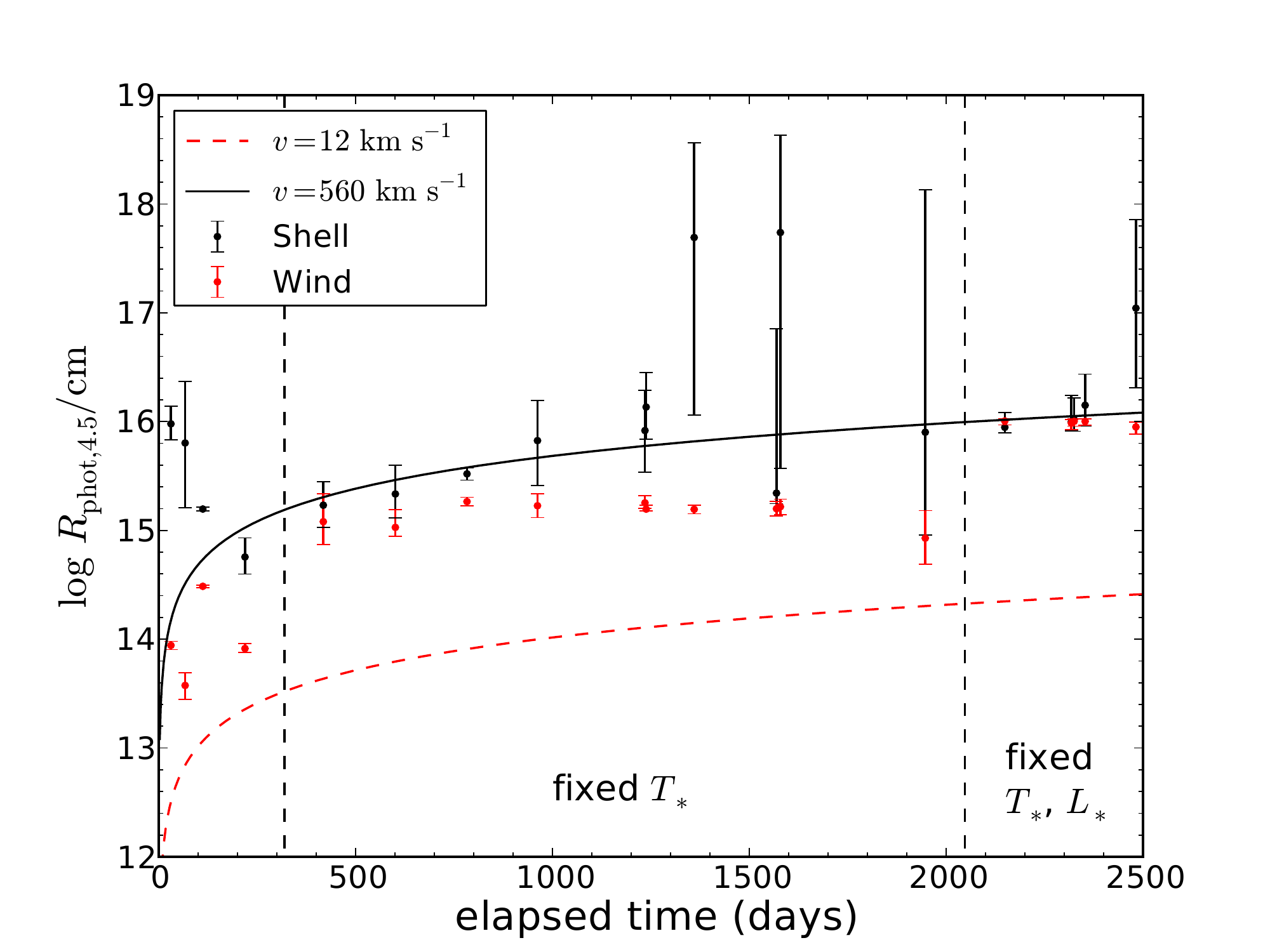}
  \fi
  \caption{Radius of the $4.5\>\mu\mathrm{m}$ photosphere, $R_{\mathrm{phot},4.5}$, of N300OT as a function of time.  The black points are the best-fit values and 1$\sigma$ uncertainties from the MCMC modeling of a dusty shell ($R_{\mathrm{out}}/R_{\mathrm{in}}=2$ with variable $T_{\mathrm{dust}}$, but without the velocity prior) for $R_{\mathrm{phot},4.5}$.  The red points are for a dusty wind ($T_{\mathrm{dust}}=1500$ K and variable $R_{\mathrm{out}}/R_{\mathrm{in}}$).  The models leftward of the first vertical, black line have variable $T_{*}$.  Due to the decreasing number of photometric constraints we fix $T_{*}=3500$ K for later epochs.  Epochs after the second vertical black line  only have detections at $4.5\>\mu\mathrm{m}$, so for these models we also fix $L_{*}$ to the progenitor luminosity. For comparison, the solid black line and the dashed red line show the radii that would be reached by material ejected at the time of the transient with the labeled velocities.  A post-outburst wind could only give rise to the inferred $R_{\mathrm{phot},4.5}$ if it has a very high velocity.\label{fig:Rphot}}
\end{figure}
% generated by: python paramevol2sets.py @names_new Rphot '$\mathrm{log}$ $R_{\mathrm{phot,4.5}}/\mathrm{cm}$' --outfile rphotnew.pdf --burn 120
% run from: /home/poseidon/sadams/impostors/dusty/ngc300ot/evolution

To illustrate this, we modelled the evolution of the $4.5\>\mu\mathrm{m}$ photosphere of N300OT over time, supplementing the photometric constraints presented in this paper with those from \citet{Bond09}, \citet{Berger09}, \citet{Prieto09}, \citet{Prieto10}, \citet{Ohsawa10}, and \citet{Hoffman11}.  In order for the post-transient obscuration to be greater than that of the progenitor and explain the decrease in the mid-IR luminosity, the density of the CSM around $R_{\mathrm{phot},4.5}$ must have increased.  However, Fig. \ref{fig:Rphot} shows that the velocity required for material ejected during or after the transient to reach $R_{\mathrm{phot},4.5}\sim10R_{\mathrm{f}}$ ($\sim 500\>\mathrm{km}\>\mathrm{s}^{-1}$) is much faster than the wind velocity of an AGB star progenitor ($\sim12\>\mathrm{km}\>\mathrm{s}^{-1}$).  Neither N300OT nor SN 2008S could have survived as cool red supergiants obscured by a new wind.

We also considered the possibility that the eruption transformed the progenitor into a hotter star with a much faster wind possibly capable of reaching $R_{4.5}$ ($\sim10R_{\mathrm{f}}$).
For $T_{*}=10000$ K (center columns of Fig. \ref{fig:fixedTs_wind}) even a fast wind ($\sim500~\mathrm{km}\>\mathrm{s}^{-1}$) with $\tau_{V,\mathrm{tot}}=1000$ that has reached $R_{\mathrm{out}}/R_{\mathrm{f}}=10$ is insufficient to hide the luminous progenitors.  Please note that the mass loss estimates shown in Figure \ref{fig:fixedTs_wind} are scaled to a wind with $10~\mathrm{km}\>\mathrm{s}^{-1}$ and should be rescaled proportionally with the wind velocity by the reader (i.e., $\times50$ for a $500~\mathrm{km}\>\mathrm{s}^{-1}$ wind reaching $10R_{\mathrm{f}}$).
For $T_{*}=30000$ K (right-hand columns of Fig. \ref{fig:fixedTs_wind}) a very fast wind ($\sim1500~\mathrm{km}\>\mathrm{s}^{-1}$) with $\tau_{V,\mathrm{tot}}=1000$ that has reached $R_{\mathrm{out}}/R_{\mathrm{f}}=30$ is able to hide the progenitor of SN 2008S but not of N300OT.
For fast winds the large optical depths require an unreasonably large mass loss of 
\begin{eqnarray}
\dot{M} \sim 0.1 \left( \frac{v_{\mathrm{w}}}{500\>\mathrm{km}\>s^{-1}} \right) \left(\frac{R_{\mathrm{f}}}{10^{15}\>\mathrm{cm}}\right) \left(\frac{\tau_{V,\mathrm{tot}}}{1000}\right) \nonumber \\ \times \left(\frac{100~\mathrm{cm}^{2}\>\mathrm{g}^{-1}}{\kappa_{V}}\right) M_{\odot}\>\mathrm{yr}^{-1} .
\end{eqnarray}
A similarly extreme mass loss rate would be required for SN 2008S to have a surviving star obscured by a dusty wind.
Although such large mass loss rates have been inferred for the progenitors of some SN IIn \citep[e.g.,][]{Chugai04,Smith10,Kiewe12} and in models \citep{Kashi15}, these were much more massive stars.

The dust that re-formed in the pre-existing wind would also contribute to the total obscuration.  
  This would essentially result in a more extended wind without requiring an increased post-eruption wind velocity.  However, as discussed in \S\ref{sec:shellresults}, much of the dust in the pre-existing wind has likely been destroyed a shock.  
Even if a shock did not sweep up the pre-existing wind and dust again formed in this CSM in addition to a heavier post-transient wind, any surviving star must be obscured by $\tau_{V,\mathrm{tot}}\sim1000$ (see $R_{\mathrm{out}}/R_{\mathrm{f}}=30$ results in Fig. \ref{fig:fixedTs_wind}).

\section{Discussion}
The main observational results of the paper are that both N300OT and SN 2008S have faded below the luminosities of their progenitors in all filters for which a comparison can be made.  We do detect $4.5~\mu\mathrm{m}$ flux at the location of these transients, but the sources are still fading.
The optical/near-IR limits and the $4.5~\mu\mathrm{m}$ detection allow a star as luminous as the progenitor to still be present, but only if the optical depth is very high.  The obscuring material must be located at large distances from the star in order to have a sufficiently cold mid-IR photosphere.  
Using a newly forming wind seems less plausible than having a dusty shell of ejecta, but the ejected mass must be $\gtrsim1~M_{\odot}$.  It is unclear whether such large, non-terminal mass ejections are plausible.
While $\eta$ Car ejected $\sim10~M_{\odot}$ during its Great Eruption \citep{Smith03}, it did so over a period of a decade or more.  Moreover, the progenitors of N300OT and SN 2008S were $<14~M_{\odot}$ \citep{Bond09,Berger09} and more likely $<10~M_{\odot}$ if extreme AGB stars \citep{Thompson09} rather than the $\sim160~M_{\odot}$ of $\eta$ Car \citep{Davidson97}.

The mass loss/ejection requirements to hide survivors to N300OT and SN 2008S would, of course, be reduced if the luminosity of the stars diminished --- the ``tuckered-out" star hypothesis \citep{Smith11}.  The `buildup' or `recovery' time-scale for the radiated energy of N300OT is
\begin{equation}
t_{\mathrm{rad}} \sim 42 \left(\frac{t_{1.5}}{80~\mathrm{days}}\right) \left(\frac{L_{\mathrm{peak}}/L_{\mathrm{_*}}}{190}\right) \zeta~ \mathrm{yr} .
\end{equation}
But as we discussed in \citet{Adams15}, this is likely not the most relevant time-scale for the stellar luminosity since the envelope will likely return to thermal equilibrium primarily through Kelvin--Helmholtz contraction rather than by energy radiated from the core.  Moreover, any transient mechanism that has no `knowledge' of the escape speed would generally leave a surviving star overexpanded and likely overluminous rather than underluminous \citep{Pan13,Shappee13}.
The birth of a massive white dwarf also seems inconsistent with the data, since the luminosity is expected to exceed pre-outburst levels for decades in this scenario \citep{Kwok93}.

\begin{figure}
  \ifpdflatex
    \includegraphics[width=8.6cm, angle=0]{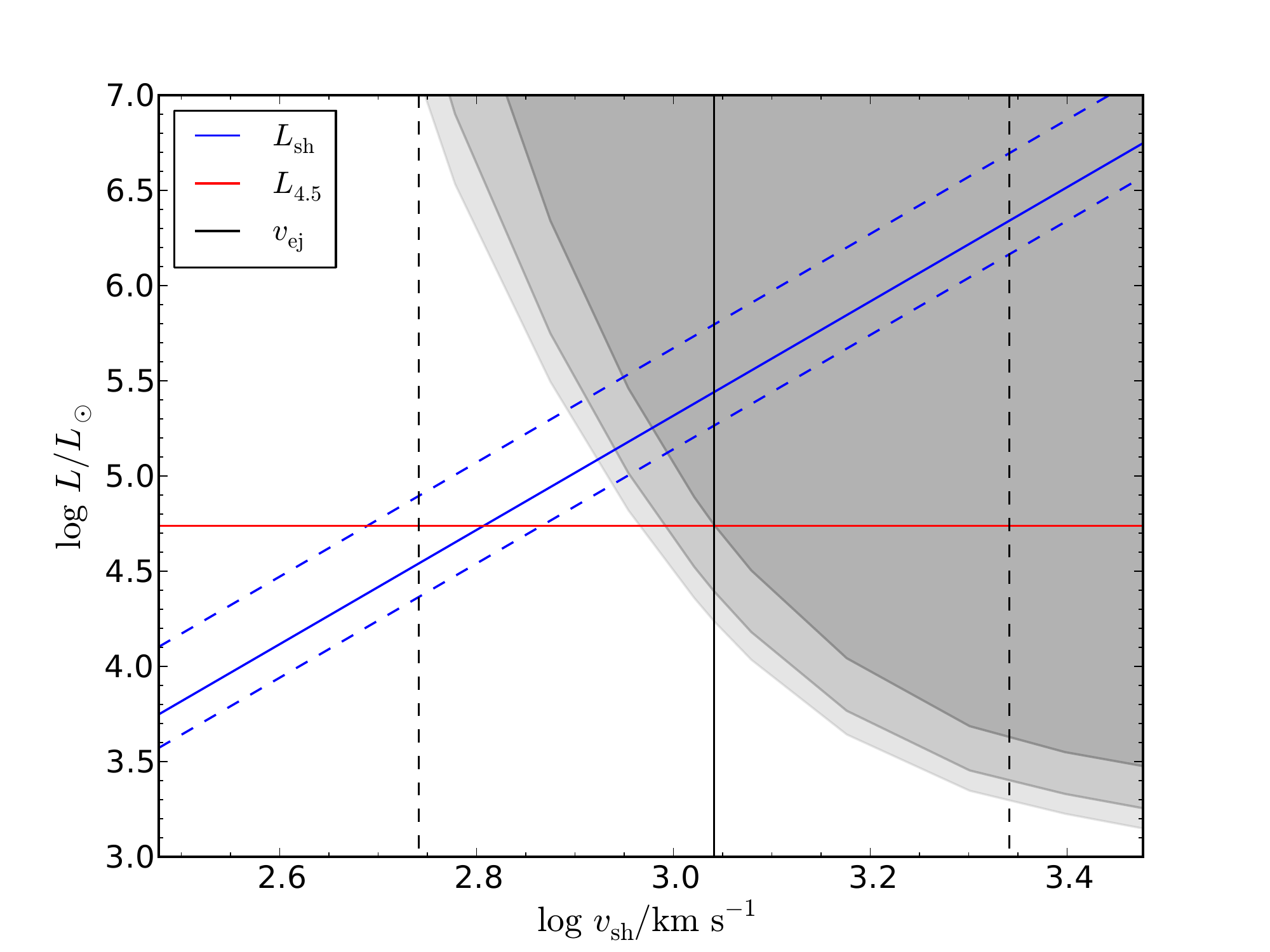}
    %\includegraphics[width=8.6cm, angle=0]{/home/poseidon/sadams/impostors/paper/300OT/shock/xraylim.pdf}
  %\else
  %  \includegraphics[width=8.6cm, angle=0]{/home/poseidon/sadams/impostors/paper/300OT/shock/xraylim.eps}
  \fi
  \caption{\emph{Chandra} constraints on the ejecta-CSM shock luminosity of SN 2008S.  The blue lines show the maximum possible luminosity of a fully radiative shock as a function of shock velocity for the best-fit wind density parameter of the progenitor (solid line) and the 1-sigma uncertainties (dashed lines).  The limit on the shock luminosity ruled out by the \emph{Chandra} non-detection on 2012-5-12 is given by the gray shaded region with the mid-tone corresponding to best-fit wind density parameter and the lighter and darker tones corresponding to the 1-sigma uncertainties.  The estimated $4.5\>\mu\mathrm{m}$ luminosity for this epoch (interpolated from the lightcurve) is shown by the red, horizontal, solid line.  The ejecta velocity and the adopted factor of two uncertainties are given by the black, vertical, solid and dashed lines.  The $4.5\>\mu\mathrm{m}$ luminosity can be powered solely by shock luminosity for shock velocities where the red line is outside of the gray shaded region and below the blue line.\label{fig:xraylim}}
\end{figure}
% figure generated by: python xraylim.py
%   run from: /home/poseidon/sadams/impostors/paper/300OT/shock

The alternative is that the transients were low-luminosity supernovae.
If the events are terminal, the fading $4.5\>\mu\mathrm{m}$ flux might be reprocessed light due to some combination of the shock luminosity from the interaction of the ejecta with the CSM \citep{Kochanek11b}, radioactivity \citep{Botticella09}, or a remnant.  With the dense winds surrounding SN 2008S and N300OT the maximum possible shock luminosity, 
\begin{eqnarray}
L_{\mathrm{s,obs}} \simeq 2.8\times10^5 \left(\frac{v_{\mathrm{e}}}{1100~\mathrm{km}\>\mathrm{s}^{-1}}\right)^{3} \nonumber \\ \times \left(\frac{\dot{M}/v_{\mathrm{w}}}{2.4\times10^{-5}~M_{\odot}\>\mathrm{yr}^{-1}/\mathrm{km}\>\mathrm{s}^{-1}}\right)~L_{\odot} ,
\end{eqnarray}
(where $\dot{M}/v_{\mathrm{w}}$ is normalized to the value for the SN 2008S progenitor found in \S2.3) is large compared to the observed IR luminosity.  
Radiating only a small fraction of this luminosity in the IR would account for the observed flux.  
The limit on the IR luminosity of SN 2008S attributable to absorbed X-ray luminosity from the forward shock found at the time of our \emph{Chandra} observation in 2012 is somewhat lower than the contemporaneous IR observations for the fiducial parameters used in Eqn. \ref{eqn:Xraylim} (see Fig. \ref{fig:xraylim}).  However, given the parameter uncertainties, the X-ray non-detection is still consistent with supernova interpretation.  For example, if the shock velocity is only $20\%$ lower than the fiducial value, an X-ray detection would not be expected even if the absorbed X-ray shock luminosity was the sole energy source for the IR luminosity observed in 2012.

Given the low masses of the progenitors and the peculiarity of the events, it has been speculated that SN 2008S-like transients could be ecSNe \citep{Botticella09}.  However, light-curve modeling of ecSNe exploding within their progenitor winds suggest that the initial transient luminosities should be significantly higher than those observed for SN 2008S and N300OT unless the envelopes were mostly lost prior to the explosions \citep{Moriya14}.  The precise mass ranges giving rise to ecSNe and ccSNe are uncertain, but \citet{Poelarends08} predict the progenitors of ecSNe to have final luminosities of $10^{5.0-5.2}\>L_{\odot}$ and the least massive ccSN progenitors to be $10^{4.6}\>L_{\odot}$.  With these estimates N300OT and SN 2008S seem more likely to be the least massive ccSNe.

Continued monitoring of SN 2008S and N300OT is needed to settle the debate on the fates of these objects.  
If the $4.5\>\mu\mathrm{m}$ fluxes continue to decrease, the mass loss or ejected mass needed to account for the obscuration will become increasingly unreasonable.  The primary problem is that without longer-wavelength data the temperature of any dust surrounding these objects is weakly constrained.  Ultimately, the issue may need to be resolved by $10-20~\mu\mathrm{m}$ observations with the \emph{James Webb Space Telescope}, which should easily determine if there is cold dust obscuring surviving stars.

\section*{Acknowledgements}
Financial support for this work was provided by NSF through grant AST-1515876 and by the National Aeronautics and Space Administration through Chandra Award Number 13099 issued by the Chandra X-ray Observatory Center, which is operated by the Smithsonian Astrophysical Observatory for and on behalf of the National Aeronautics Space Administration under contract NAS8-03060.
Support for J.L.P. is in part provided by FONDECYT through the grant 1151445 and by the Ministry of Economy, Development, and Tourism's Millennium Science Initiative through grant IC120009, awarded to The Millennium Institute of Astrophysics, MAS.  B.S. is a Hubble/Carnegie-Princeton Fellow, and supported by NASA through Hubble Fellowship grant HF-51348.001 awarded by the Space Telescope Science Institute, which is operated by the Association of Universities for Research in Astronomy, Inc., for NASA, under contract NAS 5-26555.  This work is based in part on observations made with the Spitzer Space Telescope, which is operated by the Jet Propulsion Laboratory, California Institute of Technology under a contract with NASA, and in part on observations made with the NASA/ESA \emph{Hubble Space Telescope} obtained at the Space Telescope Institute.
These observations are associated with program GO-13477.
This work is also based in part on observations made with the Large
Binocular Telescope. 
The LBT is an international collaboration among institutions in the United States, Italy and Germany. The LBT Corporation partners are: The University of Arizona on behalf of the Arizona university system; Istituto Nazionale di Astrofisica, Italy;  LBT Beteiligungsgesellschaft, Germany, representing the Max Planck Society, the Astrophysical Institute Potsdam, and Heidelberg University; The Ohio State University; The Research Corporation, on behalf of The University of Notre Dame, University of Minnesota and University of Virginia.
This paper includes data gathered with the 6.5 meter Magellan Telescopes located at Las Campanas Observatory, Chile.

\begin{appendix}
\section{Dust Models for all Stellar Temperatures}
\label{app:one}

\begin{figure*}
  \ifpdflatex
    \includegraphics[width=0.9\textwidth]{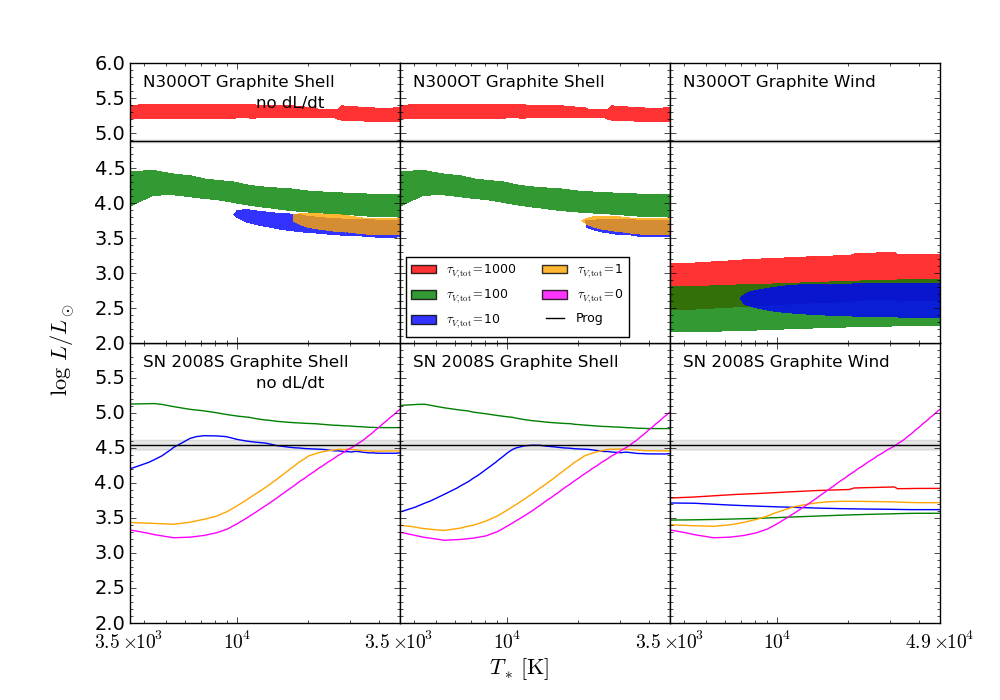}
  \else
    \includegraphics[width=0.9\textwidth]{figA1.ps}
  \fi
  \caption{SED modeling results for different obscuration scenarios, showing the possible luminosities of a surviving star as a function of stellar temperature when treating the $4.5~\mu\mathrm{m}$ flux as a detection.  The colored bands in the top panels show the luminosities within the 99.99\% confidence intervals ($\Delta \chi^2 < 21.1$ for three parameters -- $L_{*}$, $T_{*}$, and $\tau$) for $\tau_{V,\mathrm{tot}}=$ 0, 1, 3, 10, 100, and 1000.  For the N300OT $\tau_{V,\mathrm{tot}}=0$ cases the $\Delta \chi^2$ is above 21.1 for the entire parameter space shown.  In the bottom panels, the colored bands are replaced by lines showing the maximum luminosities within 99.99\% confidence intervals because the lower limits extend to a luminosity of zero (the $4.5~\mu\mathrm{m}$ flux detection is at less than $2\sigma$).  The solid black horizontal line and surrounding gray shading indicates the progenitor luminosity and $1\sigma$ uncertainties.  For these models N300OT and SN 2008S could only have survived at their pre-outburst luminosities as cool ($\sim3500$ K) super-AGB stars if they are currently obscured by dusty shells with $\tau_{V,\mathrm{tot}} > 100$ and $\tau_{V,\mathrm{tot}} > 10$, respectively.  Even if we allow the surviving stars to have become much hotter the results are essentially unchanged for N300OT while the obscuration required to hide SN 2008S is still $\tau_{V,\mathrm{tot}} > 1$.\label{fig:detectionmodels}}
\end{figure*}
% python ../../../scripts/compile_dusty_results.py from: /home/poseidon/sadams/impostors/dusty/SN2008S/arjuna
% python ../scripts/foe_figures_detection_oct.py from: /home/poseidon/sadams/impostors/dusty

\begin{figure*}
  \ifpdflatex
    \includegraphics[width=0.9\textwidth]{figA1.png}
  \else
    \includegraphics[width=0.9\textwidth]{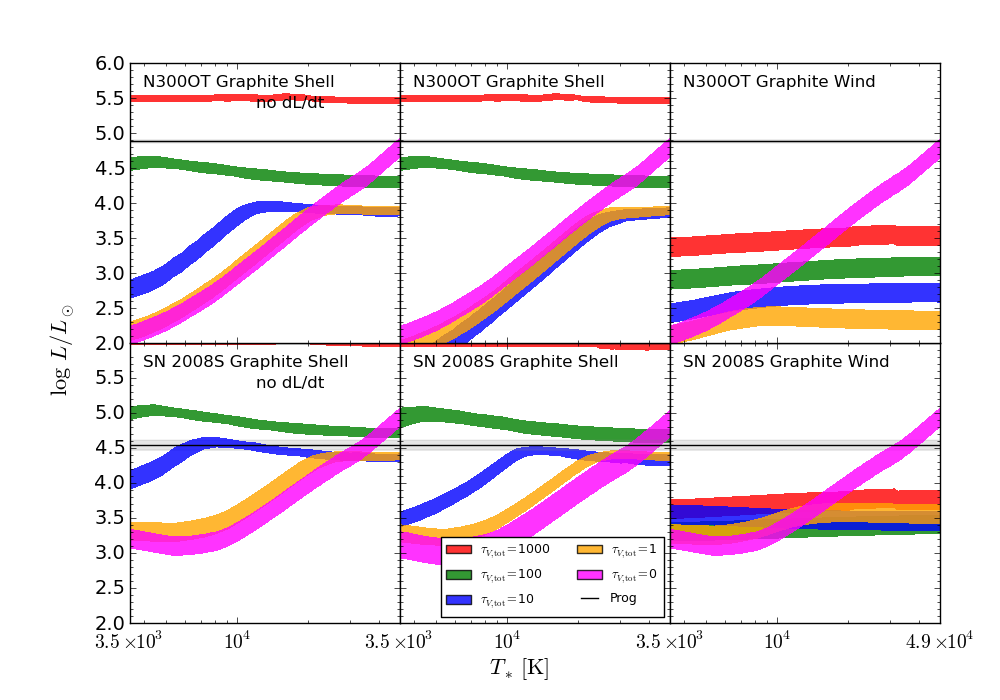}
  \fi
  \caption{SED modeling results for different obscuration scenarios, showing the possible luminosities of a surviving star as a function of stellar temperature when treating our photometric constraints as upper limits.  The colored bands show the luminosities within the 90-99.9\% confidence intervals ($6.23 < \Delta \chi^2 < 21.1$ for three parameters -- $L_{*}$, $T_{*}$, and $\tau$) relative to no surviving star for $\tau_{V,\mathrm{tot}}=$ 0, 1, 3, 10, 100, and 1000. The solid black horizontal line and surrounding gray shading indicates the progenitor luminosity and $1\sigma$ uncertainties.  For these  models N300OT and SN 2008S could only have survived at their pre-outburst luminosities as cool ($\sim3500$ K) super-AGB stars if thy are currently obscured by a dusty shell with $\tau_{V,\mathrm{tot}} > 100$ and $\tau_{V,\mathrm{tot}} > 10$.  Even if we allow the surviving stars to have become much hotter the results are essentially unchanged for N300OT while the obscuration required to hide SN 2008S is still $\tau_{V,\mathrm{tot}} > 1$.\label{fig:limitsmodels}}
\end{figure*}
% python ../../../scripts/compile_dusty_results.py from: /home/poseidon/sadams/impostors/dusty/SN2008S/arjuna
% python ../scripts/foe_figures_oct.py from: /home/poseidon/sadams/impostors/dusty

Although a surviving star would likely have a temperature similar to one of the three cases ($T_{*}=3500$, $10^{4}$, or $3\times10^{4}$ K) we have already discussed, for completeness in Figure \ref{fig:detectionmodels} we present the full grid of models we calculated for stellar temperatures ranging from 3500 to 49,000 K with $\tau_{V,\mathrm{tot}}=0$, 10, 100, and 1000.  We also show the results for the shell model when not including the variability constraints (the left-hand column of Fig. \ref{fig:detectionmodels}).
The variability limits have the largest impact at low to moderate levels of obscuration ($\tau_{V,\mathrm{tot}} = 1$ or 10) where the expanding shell model would imply the largest $dL/dt$ in optical filters, but these optical depths are already ruled out by the latest photometric constraints at all stellar temperatures for N300OT and at cool temperatures for SN 2008S.
Well-fitting models for N300OT are not possible for the $\tau_{V,\mathrm{tot}}=0$ shell model or for the $\tau_{V,\mathrm{tot}}=0$ and 1 wind models because the $4.5~\mu\mathrm{m}$ flux cannot be fit by a stellar source at low optical depth without violating the optical and near-IR upper limits.  Meanwhile, the flux for SN 2008S is only a $1\sigma$ detection.

Since the $4.5~\mu\mathrm{m}$ fluxes are still declining and might be from a
shock rather than a surviving star we also provide another
set of models that treat the $4.5~\mu\mathrm{m}$ detections as upper
limits to the luminosity from a dusty survivor (see Fig. \ref{fig:limitsmodels})
As one would expect, the constraints on the luminosity of a surviving star from the ``detection" models in Figure \ref{fig:detectionmodels} roughly define the upper limits in these models, but it is now possible to see limits on a survivor to N300OT at the lower optical depths and cooler temperatures that could not produce good fits in the ``detection" case.
Though this figure appears to show that the progenitors of N300OT and SN 2008S could have survived unobscured as hot ($T_{*}\gtrsim5\times10^{4}$ and $3\times10^{4}$ K, respectively) stars, we discussed in \S\ref{sec:shellresults} that the early optical depth evolution of the transients implies that they cannot be unobscured now.

\end{appendix}

\bibliography{references}
\bibliographystyle{mn2e}

\begin{table}
\begin{minipage}{4in}
\caption{SN 2008S Progenitor Photometry}
\begin{tabular}{lll}
\hline
\hline
{Filter} & {Magnitude$^a$} & {Luminosity$^b$ [$L_{\odot}$]} \\
\hline
U (LBT-LBC-Blue) &      $>25.8$ mag$^c$ &       $<4600$ \\
B (LBT-LBC-Blue) &      $>25.9$ mag$^c$ &       $<4200$ \\
V (LBT-LBC-Blue) &      $>26.0$ mag$^c$ &       $<1800$ \\
R (Gemini-GMOS) &       $>24.5$ mag$^d$ &       $<4300$ \\
I (Gemini-GMOS) &       $>22.9$ mag$^d$ &       $<10100$        \\
K' (MMT-PISCES) &       $>18$ mag$^e$   &       $<63300$        \\
$3.6~\mu\mathrm{m}$ (\emph{Spitzer} IRAC) & $<5~\mu\mathrm{Jy}^c$        &      $<4300$ \\
$4.5~\mu\mathrm{m}$ (\emph{Spitzer} IRAC) & $22 \pm 3 \mu\mathrm{Jy}^c$  &      $14800\pm2000$  \\
$5.8~\mu\mathrm{m}$ (\emph{Spitzer} IRAC) & $49 \pm 12 \mu\mathrm{Jy}^c$ &      $25300\pm6200$  \\
$8.0~\mu\mathrm{m}$ (\emph{Spitzer} IRAC) & $66 \pm 13 \mu\mathrm{Jy}^c$ &      $24500\pm4800$  \\
$24~\mu\mathrm{m}$ (\emph{Spitzer} MIPS)  & $<96~\mu\mathrm{Jy}^c$       &      $<11900$        \\
$70~\mu\mathrm{m}$ (\emph{Spitzer} MIPS)  & $<9340~\mu\mathrm{Jy}^c$     &      $<384000$       \\
\hline
\hline
\label{tab:prog08s}
\end{tabular}
\end{minipage}
$^a$ Apparent magnitude \\
$^b$ Luminosity after correcting for Galactic extinction \\
$^c$ \cite{Prieto08} \\
$^d$ \cite{Welch08}  \\
$^e$ \cite{Botticella09}
%\end{tabular}
%\end{minipage}
\end{table}
% flux_nu numbers converted from Prieto08's flux_lambda numbers with impostors/scripts/flam2fnu.py

\begin{table}
\begin{minipage}{4in}
\caption{N300OT Progenitor Photometry}
\begin{tabular}{lll}
\hline
\hline
{Filter} & {Magnitude$^a$} & {Luminosity$^b$ [$L_{\odot}$]} \\
\hline
\emph{HST}/ACS F475W & $>28.3$ mag &    $<14$   \\
\emph{HST}/ACS F606W & $>28.5$ mag &    $<7.6$  \\
\emph{HST}/ACS F814W & $>26.6$ mag &    $<23$   \\
% Note: Berger et al. 2009 give 5-sigma HST F475,606,814 AB magnitudes
%       as does Berger & Soderberg 2008
%       (as oppossed to the 3-sigma Vega magnitudes from Bond et al. 2009)
\emph{SST} $3.6~\mu\mathrm{m}$ & $6.7 \pm 0.7 ~\mu\mathrm{Jy}$ &        $620\pm60$      \\
\emph{SST} $4.5~\mu\mathrm{m}$ & $77 \pm 9 ~\mu\mathrm{Jy}$    &        $5600\pm600$    \\
\emph{SST} $5.8~\mu\mathrm{m}$ & $325 \pm 30 ~\mu\mathrm{Jy}$  &        $18600\pm1700$  \\
\emph{SST} $8.0~\mu\mathrm{m}$ & $877 \pm 90 ~\mu\mathrm{Jy}$  &        $36300\pm3700$  \\
\emph{SST} $24~\mu\mathrm{m}$  & $2523 \pm 250~\mu\mathrm{Jy}$ &        $35300\pm3500$  \\
\hline
\hline
\label{tab:n300prog}
\end{tabular}
\end{minipage}
$^a$ Apparent magnitude \\
$^b$ Luminosity after correcting for Galactic extinction \\
%\label{tab:NGC300progenitorphotometry}
The \emph{HST} photometry is from \citet{Berger09} and \citet{Bond09} and the \emph{SST} photometry is from \citet{Prieto08atel}.
%(ASIDE: while Berger did report first, Bond gave 3-sigma Vega upper limits (rather than 5 sigma AB limits), so Bond's numbers were just easier to use...)
\end{table}

\begin{table}
\begin{minipage}{10cm}
\caption{SN 2008S \emph{SST} Light Curve}
\begin{tabular}{llrr}
\hline
\hline
{Date (UT)} & {MJD} & {$L_{3.6\,\mu\mathrm{m}}$ $[L_{\odot}]$} & {$L_{4.5\,\mu\mathrm{m}}$ $[L_{\odot}]$} \\
\hline
2008-2-6        &       54502.8 &       $1.36(1) \times 10^{6}$ &       $1.32(1) \times 10^{6}$ \\
2008-7-18       &       54665.7 &       $2.47(3) \times 10^{5}$ &       $2.92(2) \times 10^{5}$ \\
2010-8-8        &       55416.8 &       $1.04(3) \times 10^{5}$ &       $1.64(2) \times 10^{5}$ \\
2010-12-1       &       55531.4 &       $82200 \pm 3000$        &       $1.32(2) \times 10^{5}$ \\
2011-7-27       &       55769.9 &       $41800 \pm 2900$       &       $90600 \pm 2100$       \\
2011-8-12       &       55785.9 &       $38700 \pm 2900$       &       $84900 \pm 2000$       \\
2012-3-16       &       56002.0 &                               &       $63300 \pm 2500$       \\
2012-8-21       &       56160.0 &       $10800 \pm 2900$        &       $45700 \pm 2000$        \\
2013-8-17       &       56521.1 &       $2200 \pm 2900$         &       $18300 \pm 2100$       \\
2014-1-3        &       56660.6 &                               &       $10700 \pm 2000$        \\
2014-3-26       &       56742.6 &                               &       $9600 \pm 2100$ \\
2014-8-20       &       56889.5 &       $1400 \pm 2900$         &       $11800 \pm 2000$       \\
2014-9-16       &       56916.2 &       $-500 \pm 2900$         &                               \\
2014-10-15      &       56945.6 &       $200 \pm 2900$          &                               \\
2015-1-31       &       57054.0 &       $-4200 \pm 2900$        &       $4500 \pm 2000$ \\
2015-9-2        &       57267.3 &       $-8500 \pm 2900$        &       $-1400 \pm 2100$ \\
2015-9-13       &       57278.8 &       $-7400 \pm 2900$        &       $-1800 \pm 2000$ \\
\hline
\hline
\label{tab:SN2008Ssst}
\end{tabular}
\end{minipage}
Based on image subtraction.  The absolute luminosity scale is based on aperture photometry of the stack of pre-explosion images and is corrected for Galactic extinction.  The uncertainties listed here are only for the relative flux changes and do not include the uncertainty in the zeropoint.
\end{table}
% Light curve is from image subtraction
%  absolute flux based on stack of pre-explosion images (ref_I1.fits and ref_I2.fits) with 4 pix radius aperture and 4-12 pix sky annulus
% /data/poohbah/0/ckochanek/LBTmonitor/N6946/SST/ISISsmall_adams2/ch1_cal_wsys.dat
% /data/poohbah/0/ckochanek/LBTmonitor/N6946/SST/ISISsmall_adams2/ch2_cal_wsys.dat

\begin{table}
\begin{minipage}{10cm}
\caption{N300OT \emph{SST} Light Curve}
\begin{tabular}{llrr}
\hline
\hline
{Date (UT)} & {MJD} & {$L_{3.6\,\mu\mathrm{m}}$ $[L_{\odot}]$} & {$L_{4.5\,\mu\mathrm{m}}$ $[L_{\odot}]$} \\
\hline
2009-12-21      &       55186.8 &       $2.05(2) \times 10^{5}$ &       \\
2010-7-27       &       55404.1 &       $99300 \pm 200$ &       \\
2010-8-16       &       55424.1 &       $92400 \pm 200$ &       \\
2010-8-31       &       55439.7 &       $84800 \pm 200$ &       \\
2011-9-11       &       55815.5 &       $11900 \pm 150$ &       $32300 \pm 130$ \\
2011-9-14       &       55819.0 &       $11800 \pm 130$ &       $32400 \pm 120$ \\
2012-1-14       &       55940.0 &       $6570 \pm 130$          &       $19800 \pm 120$ \\
2012-8-10       &       56149.6 &       $2110 \pm 150$          &       $8350 \pm 120$  \\
2012-8-20       &       56159.2 &       $2030 \pm 130$          &       $8220 \pm 110$  \\
2013-8-23       &       56527.2 &       $570 \pm 180$           &       $2470 \pm 130$  \\
2014-3-13       &       56729.5 &       $-80 \pm 130$           &       $980 \pm 110$           \\
2014-8-29       &       56898.3 &       $-30 \pm 130$           &       $770 \pm 110$           \\
2014-9-5        &       56905.2 &       $-70 \pm 130$           &       $850 \pm 130$           \\
2014-10-3       &       56933.4 &       $-110 \pm 130$          &       $640 \pm 110$           \\
2015-2-9        &       57062.4 &       $-10 \pm 130$           &       $140 \pm 110$           \\
\hline
\hline
\label{tab:N300OTsst}
\end{tabular}
\end{minipage}
Based on image subtraction.  The absolute luminosity scale is based on aperture photometry of the stack of pre-explosion images and is corrected for Galactic extinction.  The uncertainties listed here are only for the relative flux changes and do not include the uncertainty in the zeropoint.
\end{table}
% python lightcurve2table.py NGC300/ngc300.c1.wsys
% python lightcurve2table.py NGC300/ngc300.c2.wsys
%   from /home/poseidon/sadams/scratch

\begin{table*}
\begin{minipage}{8.5in}
\caption{SN 2008S Late-time Photometry}
\begin{tabular}{llll}
\hline
\hline
{Filter} & {Magnitude$^a$} & {Luminosity$^b$ [$L_{\odot}$]} & {Epoch} \\
\hline
% hst aperture photometry from LBTmonitor/N6946/HST/stacks/aperture_photometry/NOTES
%   the upper limits might not be done correctly:
%       I just used the aperture photometry mag when it was brighter than the limiting mag
%       I probably should do aperture mag - uncertainty (or something like that)
\emph{HST} WFC3/UVIS F438W      & $>27.03$      & $<1400$       & 2014-02-21    \\
\emph{HST} WFC3/UVIS F606W      & $>27.44$      & $<400$        & 2014-02-21    \\
\emph{HST} WFC3/UVIS F814W      & $>26.42$      & $<410$        & 2014-02-21    \\
%HST WFC3/IR F110W      & $>22.68$      & 2013-12-29    \\
%HST WFC3/IR F160W      & $>21.89$      & 2013-12-29    \\
\emph{HST} WFC3/IR F110W        & $>24.81$      & $<720$        & 2012-08-30, 2013-12-29, 2015-05-12    \\
\emph{HST} WFC3/IR F160W        & $22.70\pm0.11$ & $2200\pm200$ & 2010-08-24 \\ % see/data/poohbah/0/ckochanek/LBTmonitor/N6946/HST/ISIS_adams/NOTES
\emph{HST} WFC3/IR F160W        & $23.74\pm0.28$ & $830\pm210$ & 2011-08-07 \\ % see/data/poohbah/0/ckochanek/LBTmonitor/N6946/HST/ISIS_adams/NOTES
\emph{HST} WFC3/IR F160W        & $>23.77$      & $<810$        & 2012-08-30, 2013-12-29, 2015-05-12    \\
% SST aperture photometry from /home/poseidon/sadams/impostors/data/N6946/NOTES (uses ISISsmall -- but ISISsmall_adams gives identical results)
\emph{SST} $3.6~\mu\mathrm{m}$ & $>20.02$ ($<3\mu\mathrm{Jy}$)    & $<2400$     & 2015-09-13    \\
\emph{SST} $4.5~\mu\mathrm{m}$ & $20.9\pm0.9$ ($0.1\pm0.1\mu\mathrm{Jy}$) & $550\pm470$ & 2015-09-13        \\
\hline
\hline
\label{tab:SN2008Sphotometry}
\end{tabular}
\end{minipage}
$^a$ Apparent magnitude \\
$^b$ Luminosity after correcting for Galactic extinction \\
\end{table*}

\begin{table*}
\begin{minipage}{8.5in}
\caption{N300OT Late-time Photometry}
\begin{tabular}{llll}
\hline
\hline
{Filter} & {Magnitude$^a$} & {Luminosity$^b$ [$L_{\odot}$]} & {Epoch} \\
\hline
Magellan R & $>24.37$ & $<280$  & 2015-01-02 \\   % I need to double check this number
\emph{HST} WFC3/IR F110W & $>25.08$ & $<48$     & 2012-07-18, 2013-12-20, 2015-05-27 \\
\emph{HST} WFC3/IR F160W & $>24.51$ & $<39$     & 2012-07-18, 2013-12-20, 2015-05-27 \\
%\emph{SST} $3.6~\mu\mathrm{m}$ & $>19.96$ ($<3~\mu\mathrm{Jy}$) & 2014-08-29 \\
\emph{SST} $3.6~\mu\mathrm{m}$ & $>20.16$ ($<2~\mu\mathrm{Jy}$) & $<220$        & 2015-02-09 \\
%\emph{SST} $4.5~\mu\mathrm{m}$ & $18.30 \pm 0.10$ ($9\pm1~\mu\mathrm{Jy}$) & 2014-08-29 \\
\emph{SST} $4.5~\mu\mathrm{m}$ & $18.89 \pm 0.17$ ($5\pm1~\mu\mathrm{Jy}$) & $370\pm50$ & 2015-02-09 \\
\hline
\hline
\label{tab:NGC300photometry}
\end{tabular}
\end{minipage}
$^a$ Apparent magnitude \\
$^b$ Luminosity after correcting for Galactic extinction \\
\end{table*}
% SST photometry from: /data/poohbah/0/ckochanek/LBTmonitor/NGC300/SST/ISIS3/photometry/NOTES
% HST photometry from: /data/poohbah/3/ckochanek/LBTmonitor/NGC300/HST/stacks/phot/NOTES
%  2-pix apertures with 6-pix width sky annulus for HST
%  4-pix apertures with 8-pix width sky annulus for SST

\begin{table*}
\begin{minipage}{10cm}
\caption{SN 2008S Late-time Variability Constraints}
\begin{tabular}{llll}
\hline
\hline
{Filter} & {Variability [$L_{\odot}\>\mathrm{yr}^{-1}$]} & {Date Range} & {Number of Epochs} \\
\hline
LBT U-band              & $280\pm330$   & 2010-03-18 -- 2015-04-19      &       21      \\
LBT B-band              & $570\pm210$   & 2010-03-18 -- 2015-04-19      &       24      \\
LBT V-band              & $200\pm150$   & 2010-03-18 -- 2015-04-19      &       24      \\
LBT R-band              & $20\pm50$     & 2010-03-18 -- 2015-04-19      &       24      \\
\emph{HST} WFC/IR F110W & $144\pm100$   & 2012-08-30 -- 2015-05-12      &       3       \\      % /data/poohbah/0/ckochanek/LBTmonitor/N6946/sn2008s/plot_lc_hst_f110.py
\emph{HST} WFC/IR F160W & $150\pm550$   & 2012-08-30 -- 2015-05-12      &       3       \\      % /data/poohbah/0/ckochanek/LBTmonitor/N6946/sn2008s/plot_lc_hst.py
\hline
\hline
\label{tab:SN2008Svariability}
\end{tabular}
\end{minipage}
\end{table*}

\begin{table*}
\begin{minipage}{10cm}
\caption{N300OT Late-time Variability Constraints}
\begin{tabular}{llll}
\hline
\hline
{Filter} & {Variability [$L_{\odot}\>\mathrm{yr}^{-1}$]} & {Date Range} & {Number of Epochs} \\
\hline
Magellan R-band         & $-70\pm100$   & 2010-08-09 -- 2015-01-02 & 3 \\       % /data/poohbah/3/ckochanek/LBTmonitor/NGC300/magellan/ISIS/plot_lc.py
\emph{HST} WFC/IR F110W & $-0.9\pm5.8$  & 2012-07-18 -- 2015-05-27 & 3 \\       % /data/poohbah/3/ckochanek/LBTmonitor/NGC300/HST/ISIS_F110W/NOTES
\emph{HST} WFC/IR F160W & $-7.7\pm8.8$  & 2012-07-18 -- 2015-05-27 & 3 \\       % /data/poohbah/3/ckochanek/LBTmonitor/NGC300/HST/ISIS_adams/NOTES
\hline
\hline
\label{tab:NGC300variability}
\end{tabular}
\end{minipage}
\end{table*}

\clearpage
\end{document}